\documentclass{article}

\usepackage{amssymb}
\usepackage{graphicx}
\usepackage{epsfig}

\begin{document}

\begin{center}
{\Large Algorithm for reduction of boundary-value problems in multistep adiabatic approximation\footnote{Submitted to Mathematics and Computer in Simulation}} \\ [5mm]

{\large \it A.A. Gusev$^{a,b}$\footnote{e-mail: gooseff\@jinr.ru}, O. Chuluunbaatar$^{a}$, V.P. Gerdt$^{a}$, B.L. Markovski$^{a}$, V.V. Serov$^{c}$, S.I. Vinitsky$^{a}$}\\ [5mm]
 { \it $^{a}$Joint Institute for Nuclear Research, Dubna, Russia\\
$^{b}$Dubna International University  of  Nature,  Society  and  Man, Dubna,  Russia\\
$^{c}$Saratov State University, Saratov, Russia}\\ [5mm]

\textbf{Abstract}
\end{center}

The adiabatic approximation is well-known method for effective
study of few-body systems in solid, molecular, atomic and nuclear physics,
using the idea of separation of "fast" and "slow" variables.
The generalization of the standard adiabatic ansatz for the case 
of multi-channel wave function when all variables treated dynamically is presented.
For this reason we are introducing the step-by-step averaging methods in order to eliminate
consequently from faster to slower variables.
We present a symbolic-numerical algorithm for reduction of multistep adiabatic equations,
corresponding to the MultiStep Generalization of Kantorovich Method,
for solving multidimensional boundary-value problems by finite element method.
An application of the algorithm to calculation of the ground and first exited states
of a Helium atom is given.
%\end{abstract}


 %\end{frontmatter}
\section{Motivation}
The adiabatic approximation is well-known method
for effective study of few-body systems
in molecular, atomic and nuclear physics.
On the base of pioneering work of Born and Oppenheimer
 \cite{BO27}
the method was applied in various problems of physics,
using the idea of separation of ``fast'' $\vec x_f$ and ``slow'' $\vec x_s$
variables  \cite{BH54}
in Hamiltonian composed by fast and slow subsystems
$H(\vec x_f,\vec x_s)=H_f(\vec x_f;\vec x_s)+H_s(\vec x_s)$
with characterized frequencies $\omega_f>\omega_s$, for example
in H\'enon-Heiles model \cite{M85} or quantum dot (QD) models \cite{4202}.

Purpose of this paper is to present algorithm for generalization of
the standard adiabatic ansatz \cite{MV89,Baer06},
\begin{eqnarray}
\langle \vec x_f,\vec x_s|n_k\rangle:=\sum\nolimits_{n_{k+1}'}
\langle \vec x_f|n_{k+1}',\vec x_s\rangle
\langle \vec x_s,n_{k+1}'|n_{k}\rangle, \label{anz1}
\end{eqnarray}
for the case of multi-channel wave function
when all variables treated dynamically  \cite{DMV87} and to give a general scheme and examples
of its application to calculation
of ground and exited states of Helium atom \cite{G98,A00}.

For this reason we are
introducing the step-by-step averaging methods in order to eliminate
consequently ordered
%from faster to slower
independent variables
($\vec x=\{\vec x_f,\vec x_s\}=\{x_N\succ x_{N-1}\succ...\succ x_1\}^T\in {\bf X}={\bf X_N}\cup...\cup{\bf X_1}$
of subspace of coordinate space  ${\bf X}\subset{\bf R}^N$)
and to improve accuracy of calculations
of the parametric basis functions and corresponded matrix elements, and to have possibility
for reducing
computer resources in multi-dimension case by using in perspective the MPI technology.

We present a symbolic-numerical algorithm for reduction of multistep adiabatic equations,
corresponding to the MultiStep Generalization of Kantorovich
Method  \cite{KK64} named below as (MSGKM),
for solving multidimensional boundary-value problems with solutions subject to
corresponding boundary conditions  \cite{KANTBP}
\begin{eqnarray}
 H\psi_{n_1}-2E_{n_1}\psi_{n_1}=0.\label{anz2}
\end{eqnarray}
Here the Hamiltonian $H=\sum_{i=1}^{N} H_{N+1-i}$
of a quantum system presented by sum of parametric Hamiltonians
$H_i\equiv H_i(x_i;x_{i-1}, ... ,x_1)$
of subsystems, depending on  subset of
%``faster''
independent variable $x_i$ and
%  ``slower''
parameters $x_{i-1}, ... ,x_1$,
%that can have characterized frequencies $\omega_N>\omega_{N-1}>...>\omega_1$
and solutions satisfy to orthogonality and normalizing conditions
\begin{eqnarray}
\langle n_{1}'|n_{1}\rangle= \int_{\bf X} dx_N...dx_{1}\psi_{n_1'}^\dag(\vec x)\psi_{n_1}(\vec x)
=\delta_{n_{1}'n_{1}}.
\label{anz4} \end{eqnarray}
For solving of problem (\ref{anz2})--(\ref{anz4}) we propose multistep generalization of
the standard adiabatic ansatz  (\ref{anz1}) in the following form:
{
\begin{eqnarray}
\psi_{n_1}(\vec x)=\psi_{n_{1}}^{(1)}(x_{N}^{},...,x_{1}^{})=
\sum_{n_{2}''}\psi_{n_{2}''}^{(2)}(x_{N}^{},...,x_{2}^{};x_{1}^{})
\chi_{n_{2}''n_{1}}^{(1)}(x_{1}^{})\qquad
\label{anz3}\\  =
\sum_{n_{N}'... n_{2}'}\!\!\!\!
\psi_{n_N'}^{(N)}(x_N^{};x_{N-1}^{},...,x_1^{})
...\chi_{n_{k+1}'n_{k}'}^{(k)}(x_{k}^{};x_{k-1}^{},...,x_{1}^{})
...
\chi_{n_{3}'n_{2}'}^{(2)}(x_{2}^{};x_{1}^{})\chi_{n_{2}'n_{1}}^{(1)}(x_{1}^{})
.\nonumber \end{eqnarray}
}
Optimization of a convergence rate of the method is possible at an appropriate choice
of characterized frequencies $\omega_N>\omega_{N-1}>...>\omega_1$ of subsystems.

\section{Algorithm  MSGKM}
Below we present symbolic algorithm  MSGKM for generation
of the boundary-value problems
realizing  multistep adiabatic expansion (\ref{anz3})
in solving eigenvalue problem  (\ref{anz2})
in a symbolic form by using a Maple system.
The examples  of  different versions  of the algorithm
are given in the next sections.

\noindent\underline{\textbf{ {Algorithm} MSGKM\hspace{0.69\textwidth}}} \\
\textbf{Input}: \\
$H=\sum_{i=1}^{N} H_{N+1-i}$
is main Hamiltonian dependent on ordered variables
$\vec x=\{x_N\succ x_{N-1}\succ...\succ x_1\}^T$ decomposed to sum
of parametric Hamiltonians  $H_i\equiv H_i(x_i;x_{i-1}, ... ,x_1)$,
dependent on subset %``faster''
independent variable $x_i$
and %``slower''
parameters $x_{i-1}, ... ,x_1$;\\
$H\psi_{n_1}-2E_{n_1}\psi_{n_1}=0, \\
\langle n_{1}'|n_{1}\rangle= \int_{\bf X} dx_N...dx_{1}\psi_{n_1'}^\dag(\vec x)\psi_{n_1}(\vec x)
=\delta_{n_{1}'n_{1}}$ \\ is  main eigenvalue problem for calculation of unknowns\\
 $\psi_{n_1}\equiv|n_1\rangle\leftrightarrow\langle\vec x| n_1 \rangle
 \equiv\psi_{n_1}(\vec x)$ and $2E_{n_1}=\varepsilon_{n_1}$.

\noindent\underline{\hspace{0.99\textwidth}}

\textbf{Output}: \\
A set of Eq($k$), $k=1,...,N$, is a set of auxiliary parametric
eigenvalue problems for calculation of
$\psi_{n_k}^{(k)}\equiv\psi_{n_{k}}^{(k)}(x_N,...,x_k;x_{k-1},...,x_1)$
and
$
\varepsilon^{(k)}_{n_k}\equiv\varepsilon^{(k)}_{n_k}(x_{k-1}...x_1)$,
where
$
\psi_{n_1}=\psi_{n_1}^{(1)}$ and $2E_{n_1}=\varepsilon^{(1)}_{n_1}$
are solutions of the main eigenvalue problem.

\noindent\underline{\hspace{0.99\textwidth}}

\textbf{Local}: \\
$\psi_{n_k}^{(k)}\equiv\psi_{n_{k}}^{(k)}(x_N,...,x_k;x_{k-1},...,x_1)$
and $\varepsilon_{n_k}\equiv\varepsilon^{(k)}_{n_k}
\equiv\varepsilon^{(k)}_{n_k}(x_{k-1}...x_1)$
are solutions of the auxiliary parametric eigenvalue problems: \\
$(\sum_{i=N+1-k}^{N} H_{N+1-i})\psi_{n_k}^{(k)}-\varepsilon_{n_k}^{(k)}\psi_{n_k}^{(k)}=0$,\\
$\langle n_{k}'|n_{k}\rangle=\int_{{\bf X_N}\cup...\cup{\bf X_{N+1-k}}}
dx_N...dx_{N+1-k}{\psi_{n_k'}^{(k)}}^\dag\psi_{n_k}^{(k)}=\delta_{n_{k}'n_{k}}$;
\\
$\langle n_{k+1}'|n_{k}\rangle\equiv\chi_{n_{k+1}'n_k}^{(k)}(x_k;x_{k-1},...,x_1)$
are auxiliary parametric solutions
defined as:\\
$\langle n_{k+1}'|n_{k}\rangle =\int_{{\bf X_N}\cup...\cup{\bf X_{k+1}}} dx_N...dx_{k+1}
{\psi_{n_{k+1}'}^{(k+1)}}^{\dag}
\psi_{n_{k}}^{(k)},
$\\
the square brackets $[,]$ means a commutator: \\$\langle n_{k+1}|
\Bigl[H_k,n_{k+1}'\rangle\Bigr]\!=\!\langle n_{k+1}|H_k n_{k+1}'\rangle\!
-\!\langle n_{k+1}|n_{k+1}'\rangle H_k$.

\noindent\underline{\hspace{0.99\textwidth}}\\
\textbf{1}: Eq($N$):=$ \{ H_{n_N}|{n_N}\rangle-\varepsilon_{n_N}|{n_N}\rangle=0,
\phantom{aaaa} \langle\psi_{n_N}^ {(N)\dag} |\psi_{n_N'}^{(N)}\rangle=\delta_{n_Nn_N'}\} $\\
\textbf{2}: Eq($N$) $\mathbf{\to} \{|{n_N}\rangle,\varepsilon_{n_N}\}$\\
\textbf{3}: \textbf{for} k:=N-1:1 step -1\\
\textbf{4}: \phantom{a} Eq($k$):=$\{(H_k+\varepsilon^{(k+1)}_{n_{k+1}}
- \varepsilon^{(k)}_{n_{k}}) \langle n_{k+1}|n_{k}\rangle
 $ \\  \phantom{4}$~~~~~~~~~~~~~+ \sum\nolimits_{n_{k+1}'} {\langle n_{k+1}|
 \Bigl[H_k,n_{k+1}'\rangle\Bigr]}\langle n_{k+1}'|n_{k}\rangle\!=\!0\}$.\\
\textbf{5}: \phantom{a} Eq($k$) $\mathbf{\to} \{\langle n_{k+1}'|n_{k}\rangle,
\varepsilon_{n_k}^{(k)}\}$\\
\textbf{6}: \phantom{a} $|n_k\rangle:=\sum\nolimits_{n_{k+1}'}|n_{k+1}'\rangle
\langle n_{k+1}'|n_{k}\rangle$  \\
\textbf{7}: \textbf{end for} \\
\textbf{8}: $
\psi_{n_1}=|n_1\rangle$, $2E_{n_1}=\varepsilon^{(1)}_{n_1}$

\noindent\underline{\hspace{0.99\textwidth}}

\section{Statement of the problem for a Helium atom ($N=3$)}

The Schr\"odinger equation for a Helium atom with total zero-angular momentum
in hyperspherical coordinates  \cite{A00}:
$\theta\equiv x_3\in{\bf X_3}=[0,\pi],\alpha\equiv x_2\in{\bf X_2}=[0,\pi],
R \equiv x_1\in{\bf X_1}=[0,+\infty)$,
 $\vec{x}=\{x_3\succ x_{2}\succ x_1\}^T\in {\bf X}={\bf X_3}\cup{\bf X_2}\cup{\bf X_1}$
reads as,
\begin{eqnarray*}
(H_3(x_3;x_2,x_1)+H_2(x_2;x_1)+H_1(x_1)-2E_i)\Psi_i(x_3,x_2,x_1)=0.
\end{eqnarray*}
Here the Hamiltonians $H_i$ of subsystems  consistent of
differential operators by independent variables and multiplication operators  of
the Coulomb potential energy $\hat V_i$ of the three interacted particles
with charges $Z_a=-1$, $Z_b=-1$, $Z_c=2$, including appropriate
choice of weight factors:
\begin{eqnarray*}
H_3(x_3;x_2,x_1)=\frac{4}{x_1^2\sin^2x_2}\hat H_3(x_3;x_2,x_1),
\\
 \hat H_3(x_3;x_2,x_1)\!=\!-\!\frac{1}{\sin x_3}\frac{\partial}{\partial x_3}\sin x_3
 \frac{\partial}{\partial x_3}
+\hat V_3(x_3;x_2,x_1),
\\
\hat V_3(x_3;x_2,x_1)=\frac{x_1\sin^2x_2}{2}\frac{Z_aZ_b}{\sqrt{1-\sin x_2\cos x_3}},
\\
H_2(x_2;x_1)=\frac{4}{x_1^2}\hat H_2(x_2;x_1),
\\
\hat H_2(x_2;x_1)=-\frac{1}{\sin^2x_2}\frac{\partial }{\partial x_2}\sin^2x_2
\frac{\partial}{\partial x_2}
+1+\hat V_2(x_2;x_1),
\\
\hat V_2(x_2;x_1)=\frac{x_1}{2}\left(\frac{Z_aZ_c}{\sin \frac{x_2}{2}  }
+\frac{Z_bZ_c}{\cos \frac{x_2}{2}}\right),
\\
H_1(x_1)=\hat H_1(x_1),\\
\hat H_1(x_1)=-\frac{1}{x_1^5}\frac{\partial }{\partial x_1}x_1^5
\frac{\partial}{\partial x_1}-\frac{4}{x_1^2}.
\end{eqnarray*}
Solutions of discrete spectrum satisfy to orthonormalization  conditions
\begin{eqnarray*}
\frac{1}{8}\int_{\bf X} \sin x_3dx_3\sin^2x_2dx_2x_1^5dx_1 \Psi_i(x_3,x_2,x_1)\Psi_j(x_3,x_2,x_1)
=\delta_{ij}
\end{eqnarray*}
and subject to the boundary conditions
\begin{eqnarray}\label{x5}
 \lim_{x_1\to 0}x_1^5\frac{\partial \Psi_i(x_3,x_2,x_1)}{\partial x_1}=0
, \quad \lim_{x_1\to \infty}x_1^5 \Psi_i(x_3,x_2,x_1)=0, \\
\lim_{x_2\to 0,\pi}\sin^2x_2\frac{\partial \Psi_i(x_3,x_2,x_1)}{\partial x_2}=0, \quad
\lim_{x_3\to 0,\pi}\sin x_3\frac{\partial \Psi_i(x_3,x_2,x_1)}{\partial x_3}=0. \nonumber
\end{eqnarray}

\section{ Algorithm 1. Example of the conventional Kantorovich method.}

We consider one-parametric boundary-value problem
with respect to fast $\vec x_f=\{x_3,x_2\}$ independent variables
\begin{eqnarray}
 \label{mu2a}
 (\frac{\hat H_3(x_3;x_2,x_1)}{\sin^2x_2}+\hat H_2(x_2;x_1)-\frac12E_{i_2}^{(2)}(x_1))
 \Psi_{i_2}^{(2)}(x_3,x_2;x_1)=0,
\\ \label{mu2ab}
\int_{{\bf X_3}\cup{\bf X_2}} \sin x_3dx_3\sin^2x_2dx_2 \Psi_{i_2}^{(2)}(x_3,x_2;x_1)\Psi_{j_2}^{(2)}(x_3,x_2;x_1)
=\delta_{i_2j_2},
 \end{eqnarray}
 and   conventional one by   independent variables
 $\vec x=\{x_3\succ x_2\succ x_1\}$,
{ \begin{eqnarray} \label{mu1a}\!\!\!\!\!\!
 \left( \frac{4\hat H_3(x_3;x_2,x_1)}{x_1^2\sin^2x_2}
 +\frac{4\hat H_2(x_2;x_1)}{x_1^2}+\hat H_1(x_1)
 -2E_{i_1}^{(1)}\right)\Psi_{i_1}^{(1)}(x_3,x_2,x_1)=0,
\\  \label{mu1ab}
\frac{1}{8}\int_{{\bf X}} \sin x_3dx_3\sin^2x_2dx_2x_1^5dx_1 \Psi_{i_1}^{(1)}(x_3,x_2,x_1)
\Psi_{j_1}^{(1)}(x_3,x_2,x_1)=\delta_{i_1j_1},
\end{eqnarray}
}
with boundary conditions following from (\ref{x5}).

In \textbf{Step 1}
we find the required solution of the problem (\ref{mu2a})
in the series expansion over the Legendre polynomials $P_{i_1}^{}(\cos x_3)$
for each values of $x_1$:
\begin{eqnarray}\label{mu23a}
\Psi_{i_2}^{(2)}(x_3,x_2;x_1)=\sum_{i_1=1}^{i_1^{\max}} P_{i_1}^{}(\cos x_3)
\chi_{i_1i_2}^{(2)}(x_2;x_1).
\end{eqnarray}
Substituting expansion (\ref{mu23a}) into equation (\ref{mu2a}) and projecting with
account of orthonormalization conditions of Legendre polynomials,
we arrive to the one-parametric problem for unknown vector eigenfunctions, $\chi_{j_1i_2}^{(2)}(x_2;x_1)$,
and corresponded eigenvalue (potential curve), $ {E_{i_2}^{(2)}(x_1)}$,
\begin{eqnarray} \nonumber
\left(
-\frac{1}{\sin^2x_2}\frac{\partial }{\partial x_2}\sin^2x_2\frac{\partial}{\partial x_2}
+1+{\hat V_2(x_2,x_1)}+\frac{i_1(i_1+1)}{\sin^2x_2}\right.
\\
\left.-\frac12{E_{i_2}^{(2)}(x_1)}\right)\chi_{i_1i_2}^{(2)}(x_2;x_1)+ \frac{1}{\sin^2x_2}\sum_{j_1=1}^{i_1^{\max}}
\hat V_{i_1j_1}^{(3)}(x_2;x_1)\chi_{j_1i_2}^{(2)}(x_2;x_1)
  =0,\label{pa3}
\\\nonumber
\hat V_{i_1j_1}^{(3)}(x_2;x_1)=
\int_{{\bf X_3}}\sin x_3dx_3
P_{i_1}(\cos x_3)\hat V_3(x_3,x_2;x_1)P_{j_1}(\cos x_3)
\end{eqnarray}
with boundary conditions following from (\ref{x5}).

Substituting expansion (\ref{mu23a}) into (\ref{mu2ab}), we have orthonormation conditions
\begin{eqnarray}\label{pa4}
\sum_{i_2=1}^{i^{\max}_2}\int_{{\bf X_2}} \sin^2x_2dx_2\chi_{i_1i_2}^{(2)}(x_2;x_1)
\chi_{j_1i_2}^{(2)}(x_2;x_1)=\delta_{i_1j_1}.
\end{eqnarray}
This one-parametric problem is solved with help of the adaptation of KANTBP program  \cite{KANTBP},
named here as  KANTBP 3.0.

\begin{figure}[t]
\includegraphics[width=0.49\textwidth,height=0.32\textwidth]{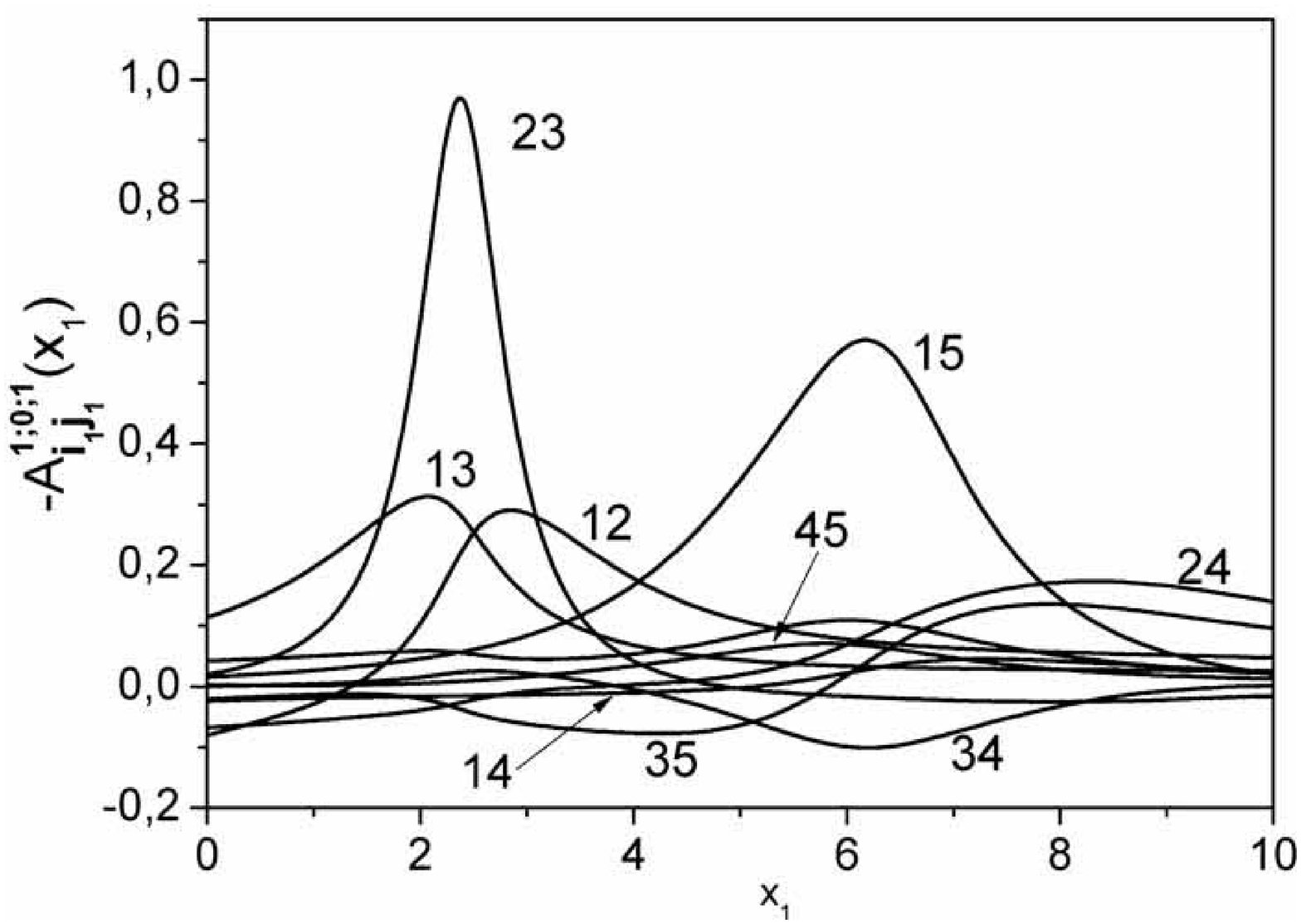}
\includegraphics[width=0.49\textwidth,height=0.32\textwidth]{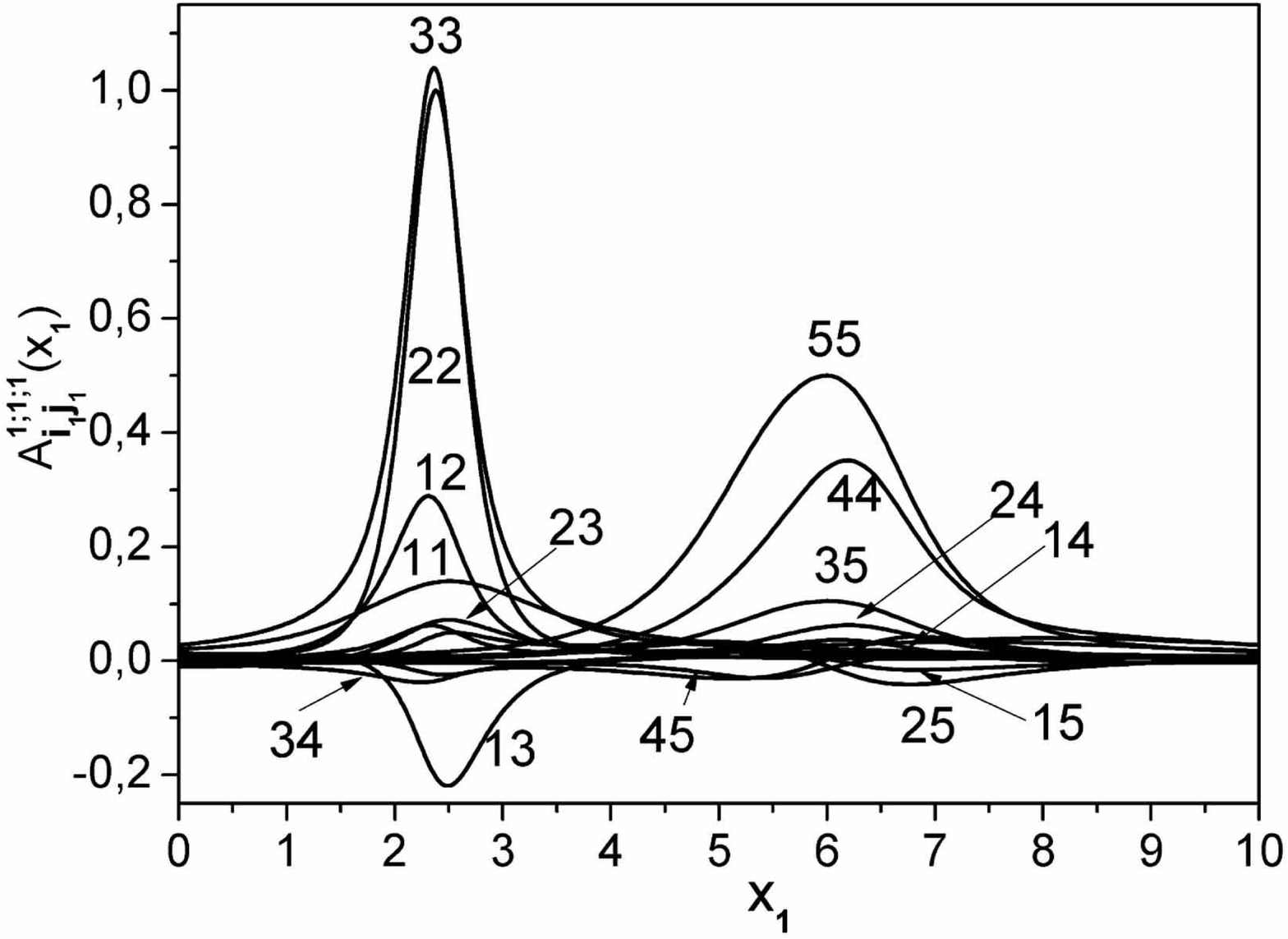}
\caption{Calculated  matrix elements in step 1 of equation in step 2. }\label{a1}
\end{figure}

\begin{figure}[t]
 \includegraphics[width=0.49\textwidth,height=0.32\textwidth]{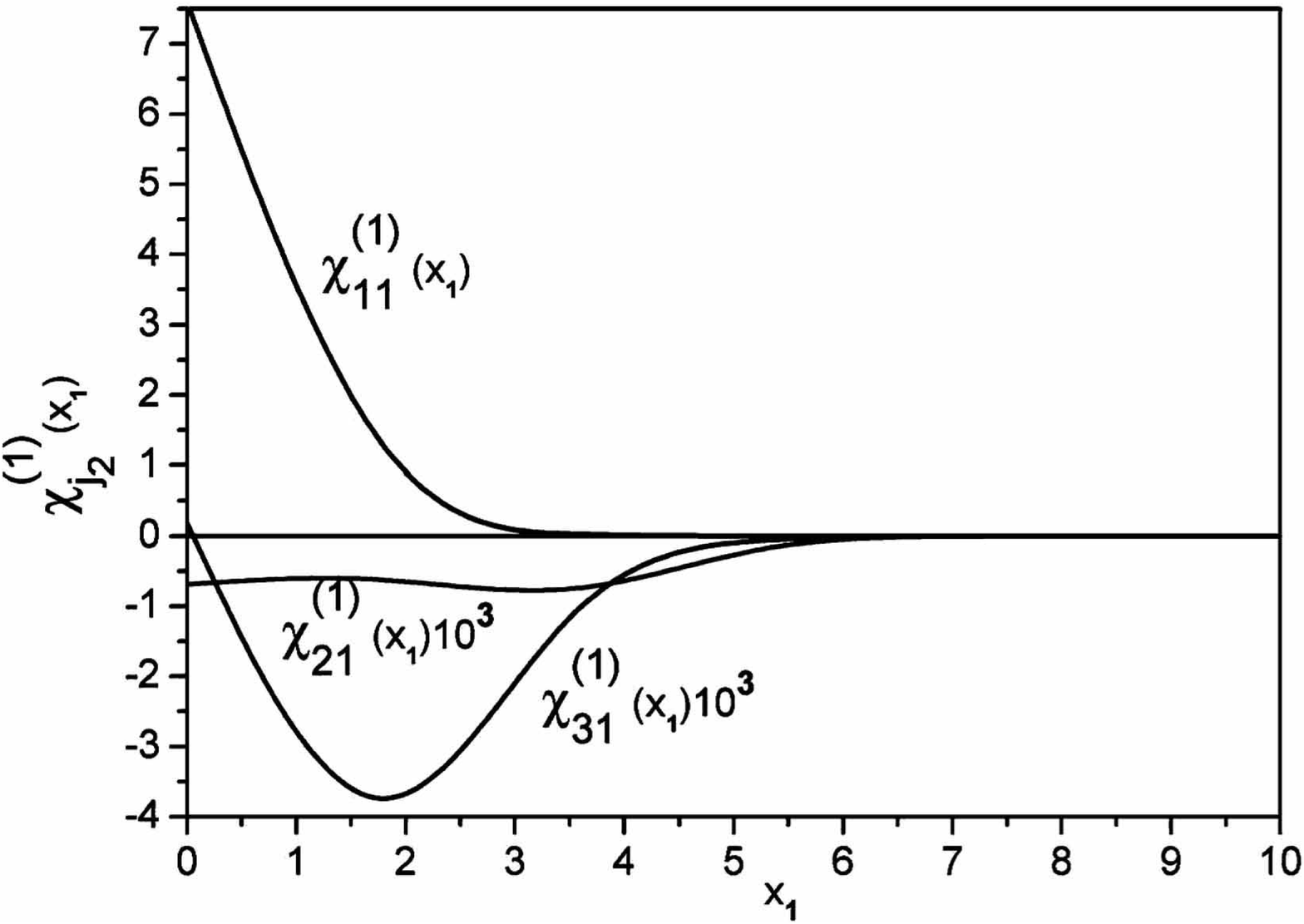}
 \includegraphics[width=0.49\textwidth,height=0.32\textwidth]{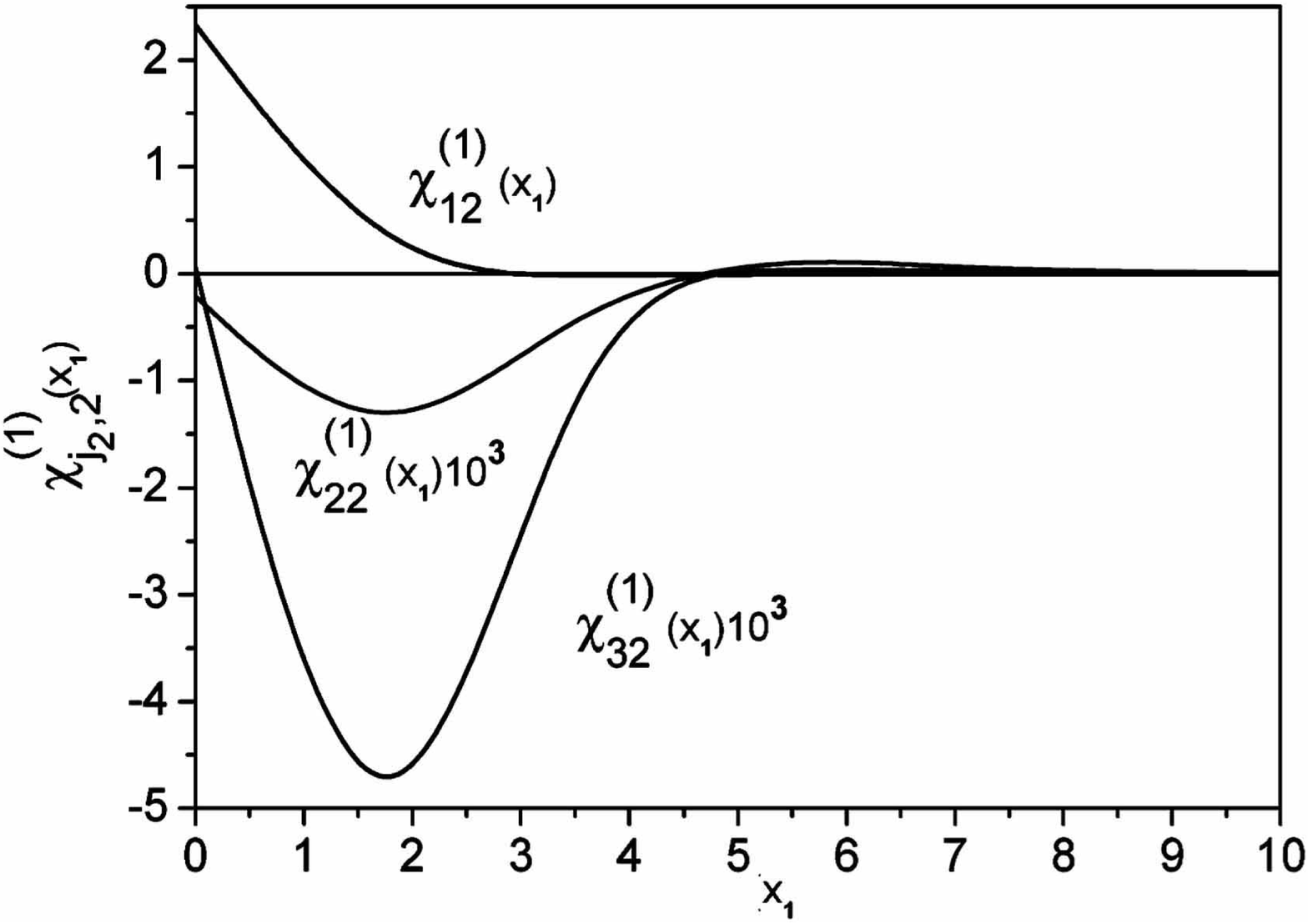}
\caption{Radial eigenfunctions of ground and first exited states.} \label{c}
\end{figure}

In \textbf{Step 2}
we find the solution of the problem (\ref{mu1a}) in the series expansion over solutions
of problem (\ref{mu2a}) solved in the \textbf{Step 1},
 \begin{eqnarray}\label{mu12a}
\Psi_{i_1}^{(1)}(x_3,x_2,x_1)=\sum_{i_2=1}^{i_2^{\max}}
\Psi_{i_2}^{(2)}(x_3,x_2;x_1)\chi_{i_2i_1}^{(1)}(x_1).
\end{eqnarray}
Substituting expansion (\ref{mu12a}) into equation (\ref{mu1a}) and projecting with
account of orthonormalization conditions  (\ref{mu2ab})
of parametric basis functions $\Psi_{i_2}^{(2)}(x_3,x_2;x_1)$ from (\ref{mu23a})
calculated in \textbf{Step 1},
we arrive to the problem for unknown vector functions, $\chi_{i_2i_1}^{(1)}(x_1)$,
and corresponding eigenenergy, $E_{i_1}^{(1)}$,
 \begin{eqnarray}
\left(
-\frac{1}{x_1^5}\frac{\partial }{\partial x_1}x_1^5\frac{\partial}{\partial x_1}
+\frac{2E_{i_2}^{(2)}(x_1)-4}{x_1^2}-2E_{i_1}^{(1)}\right)\chi_{i_2i_1}^{(1)}(x_1) \nonumber
\\ +\sum_{j_2=1}^{i^{\max}_2}
 \langle i_2|\Bigl[H_1,j_2\rangle\Bigr]\chi_{j_2i_1}^{(1)}(x_1)=0, \nonumber
\\
 {\langle i_2|\Bigl[H_1,j_2\rangle\Bigr]}=\left(A^{1;1;1}_{i_2j_2}(x_1)
-\frac{1}{x_1^5}\frac{\partial}{\partial x_1}x_1^5A^{1;0;1}_{i_2j_2}(x_1)
- A^{1;0;1}_{i_2j_2}(x_1)\frac{\partial}{\partial x_1} \right)\label{pa1}
\end{eqnarray}
with boundary conditions following from (\ref{x5}).
\begin{figure}[t]
\includegraphics[width=0.49\textwidth,height=0.32\textwidth]{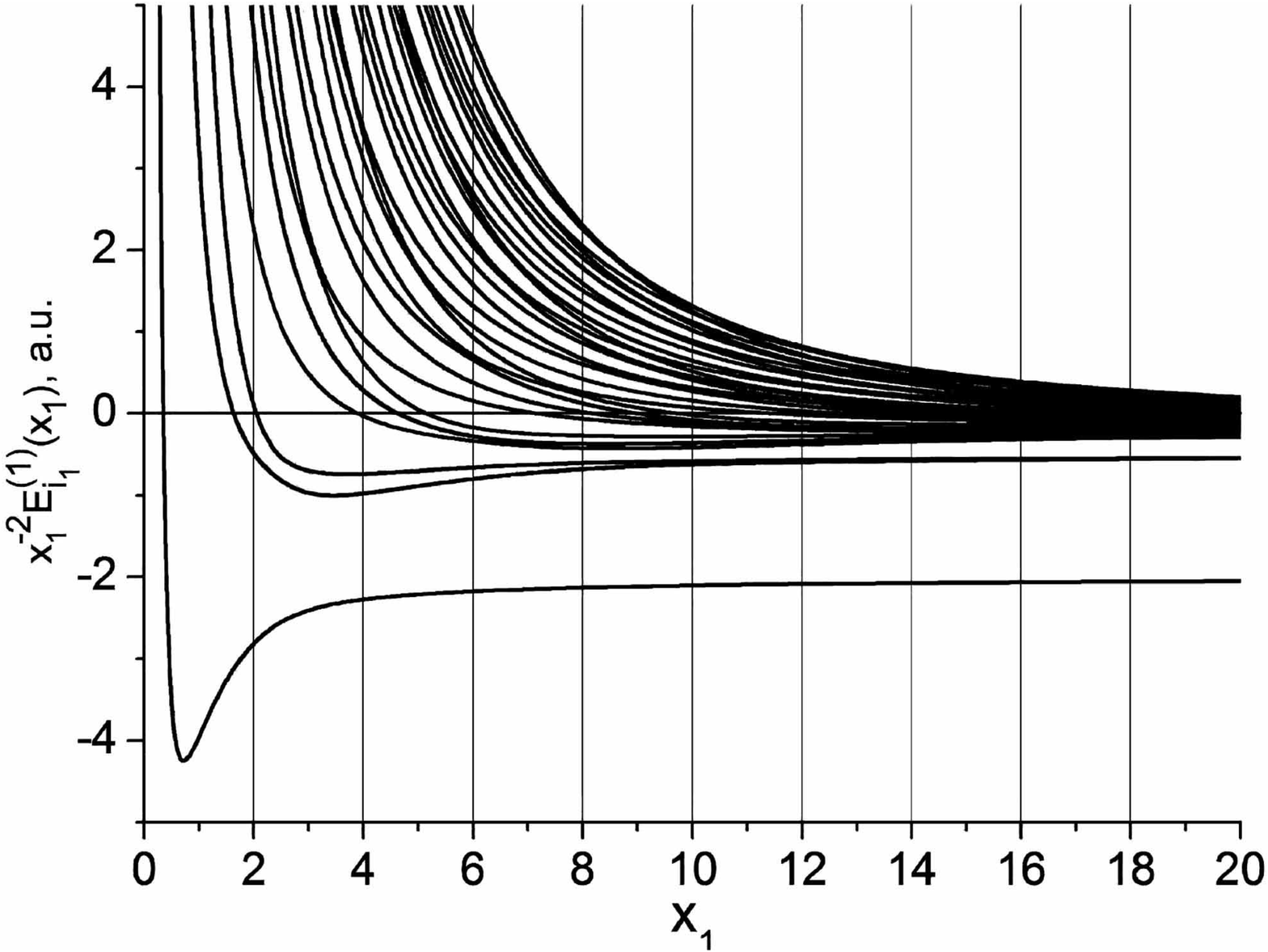}
\includegraphics[width=0.49\textwidth,height=0.32\textwidth]{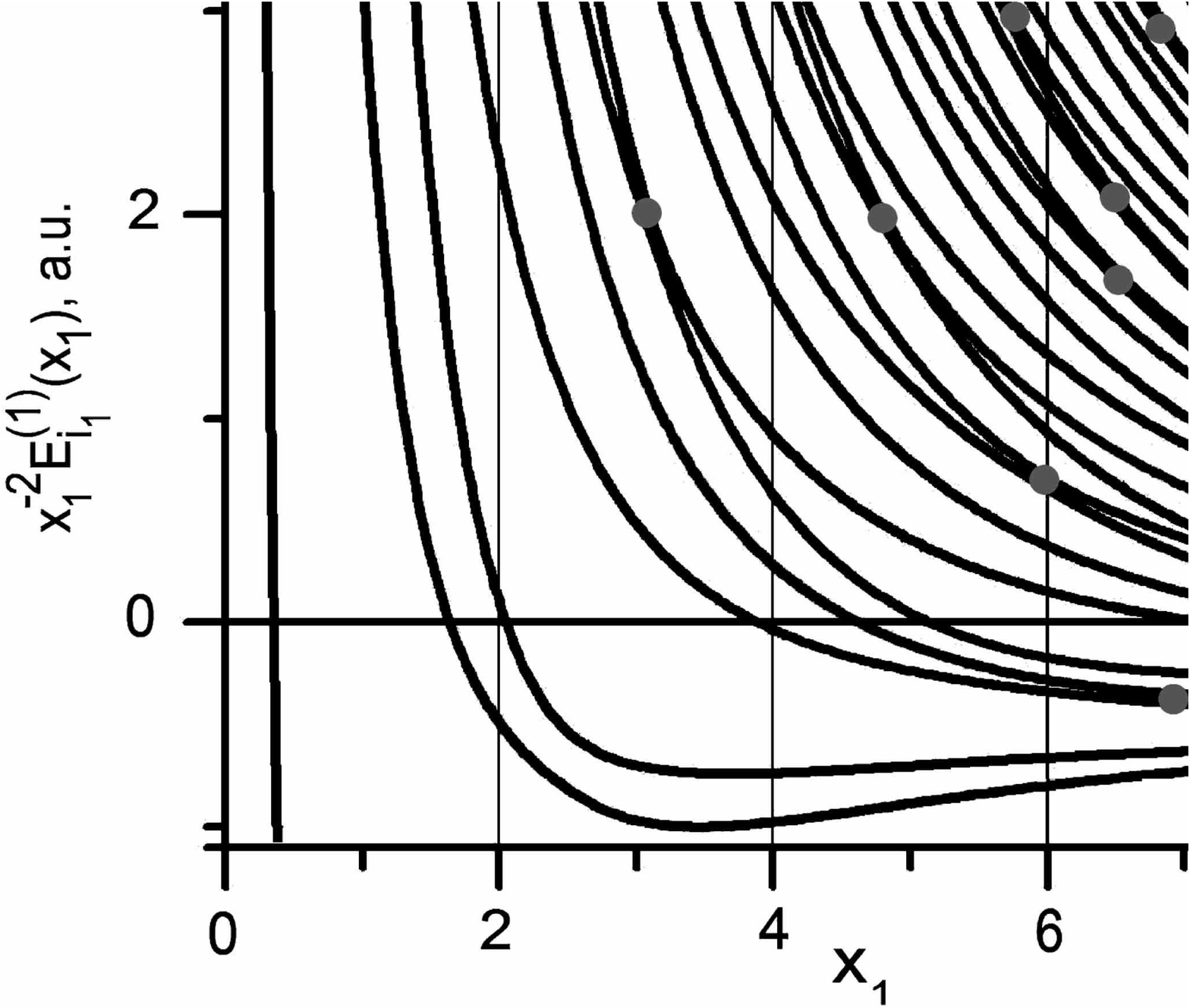}
\caption{Calculated  potential curves in step 1 of equation in step 2.
Circles on right panel note avoiding crossing points. }\label{a0}
\end{figure}
Substituting expansion (\ref{mu12a}) into (\ref{mu1ab}), we have orthonormation conditions
 \begin{eqnarray}
\sum_{j_2=1}^{i^{\max}_2}\frac{1}{8}\int_{{\bf X_1}} x_1^5dx_1\chi_{j_2i_1}^{(1)}(x_1)
\chi_{j_2j_1}^{(1)}(x_1)=\delta_{i_1j_1}.\label{pa2}
\end{eqnarray}
In (\ref{pa1}) we have definitions of elements of matrix of effective potentials ($l_1=0,1$):
{  \begin{eqnarray}
A^{1;l_1;r_1}_{i_2j_2}(x_1)\!=\!\!
\int_{{\bf X_3}\cup{\bf X_2}}\!\!  \sin x_3dx_3\sin^2x_2dx_2
\frac{\partial^{l_1}\Psi_{i_2}^{(2)}(x_3,x_2;x_1)}{\partial x_1^{l_1}}
\frac{\partial^{r_1}\Psi_{j_2}^{(2)}(x_3,x_2;x_1)}{\partial x_1^{r_1}},\!\!\!\!\!\!\!\nonumber\\
\frac{\partial^{0}}{\partial x_1^{0}\Psi_{i_2}^{(2)}(x_3,x_2;x_1)}
\equiv\Psi_{i_2}^{(2)}(x_3,x_2;x_1).\qquad\label{pd}
\end{eqnarray}}
A parametric derivatives of eigenfunction $\Psi_{j_2}^{(2)}(x_3,x_2;x_1)$
in (\ref{pd}) are calculated with help of KANTBP 3.0
program together with corresponding integrals, where integration by variable
$x_3$ perform analytically by using of orthonormalization
conditions of Legendre polynomials.
As an example, some potential curves and matrix elements of effective potential matrix
are shown in Figs. \ref{a0} and \ref{a1}.
One can see that these matrix elements have smooth behavior with respect to parameter $x_1$
that achieve by imposing conditions of continuity of eigenfunctions with respect
to parameter $x_1$ in avoiding crossing points of potential curves $E_{i_2}^{(2)}(x_1)$,
shown by circles in Fig. \ref{a0},
where change number of zeros of corresponding pair of eigenfunctions by each
of independent variables $x_2$ and $x_3$ occurs after passing these points,
as discussed in \cite{A00,AAPV91}.
As we show in Table \ref{tab}, such continuation via these points
leads to increase of the convergence rate of expansion (\ref{mu12a}) of total solution
in calculation of ground and first exited states  energy  $E^{(1)}_{i_1}$
with respect to previous one \cite{A00}.
\begin{table}[t]
\caption{Ground state 1s1s energy $E^{(1)}_2$ and
first exited state 1s2s energy  $E^{(1)}_2$ of Helium atom (in a.u.) versus number $i_2^{\max}$
of basis functions and number $i_1^{\max}$ of the Legengre polynomials}\label{tab}
\begin{tabular}{|r|ccc|l|}
\hline
$i_2^{\max}$&  1s1s: $i_1^{\max}=12$&  1s1s: $i_1^{\max}=21$& 1s1s: $i_1^{\max}=28$
&1s2s: $i_1^{\max}=28$\\
 \hline
 1 & $-$2.895 539 01 &  $-$2.895 551 19 & $-$2.895 552 76 &  $-$2.139 935 68             \\
 2 & $-$2.898 631 39 &  $-$2.898 643 21 & $-$2.898 644 74 &  $-$2.141 664 33             \\
 6 & $-$2.903 643 86 &  $-$2.903 655 95 & $-$2.903 657 51 &  $-$2.145 700 22             \\
10 & $-$2.903 702 68 &  $-$2.903 714 86 & $-$2.903 716 36 &  $-$2.145 915 09             \\
15 & $-$2.903 708 49 &  $-$2.903 720 68 & $-$2.903 722 17 &  $-$2.145 957 35             \\
\hline
21 & $-$2.903 709 31 &  $-$2.903 721 50 & $-$2.903 722 994&  $-$2.145 968 77\\
28 & $-$2.903 709 31 &                  & $-$2.903 722 997&  $-$2.145 970 28\\\hline\hline
 \cite{G98}
   &                 &                & $-$2.903 722 998&$-$2.145 956 975\\
 \cite{D94}
   &                 &                & $-$2.903 724 377& $-$2.145 974 046  \\ \hline
\end{tabular}
\end{table}
One can see from the Table \ref{tab} that convergence  start from $i_2^{\max}=21$
is slow with respect to upper variational estimation \cite{D94}.
So, to improve convergence of calculation
of the parametric basis functions from expansion (\ref{mu23a}) by number $i_1^{\max}>28$,
we begin to study in the next section the step-by-step averaging method for realizing calculation
with a more high accuracy with help of Algorithm 2.
Meanwhile, our upper estimation at $i_2^{\max}=28$ for first exited state
is lowing than result of  \cite{G98}. On Fig. \ref{c}
we show the radial eigenfunctions of ground and first exited states. Note that, as following
from asymptotic effective potentials (see Fig. \ref{a1})
the ground state solution has asymptotic in a vicinity triple-collision point $R\to0$
including logarithmic terms that corresponding to Fock expansion  \cite{AAPV91,KV87}.

\section{
Algorithm 2. Example of MultiStep Generalization of Kantorovich Method (MSGKM)}

We examine a split sequence of boundary-value problems
consists of
the two-parametric  problem by one of fast independent variables, $\vec x_f=\{x_3\}$,
\begin{eqnarray}  \label{mu3}
(\hat H_3(x_3;x_2,x_1)-\frac12E_{i_3}^{(3)}(x_2,x_1))\Psi_{i_3}^{(3)}(x_3;x_2,x_1)=0,
\\ \nonumber
\int_{{\bf X_3}}\sin x_3 dx_3 \Psi_{i_3}^{(3)}(x_3;x_2,x_1)\Psi_{j_3}^{(3)}(x_3;x_2,x_1)
=\delta_{i_3j_3},
\end{eqnarray}
the one-parametric problem (\ref{mu2a})--(\ref{mu2ab})  by fast independent variables $\vec x_f=\{x_3\succ x_2\}$
and conventional problem (\ref{mu1a})--(\ref{mu1ab}) by
    independent variables    $\vec x=\{x_3\succ x_2\succ x_1\}$
with corresponding boundary conditions following from (\ref{x5}).

\begin{figure}[t]
\includegraphics[width=0.49\textwidth,height=0.23\textwidth]{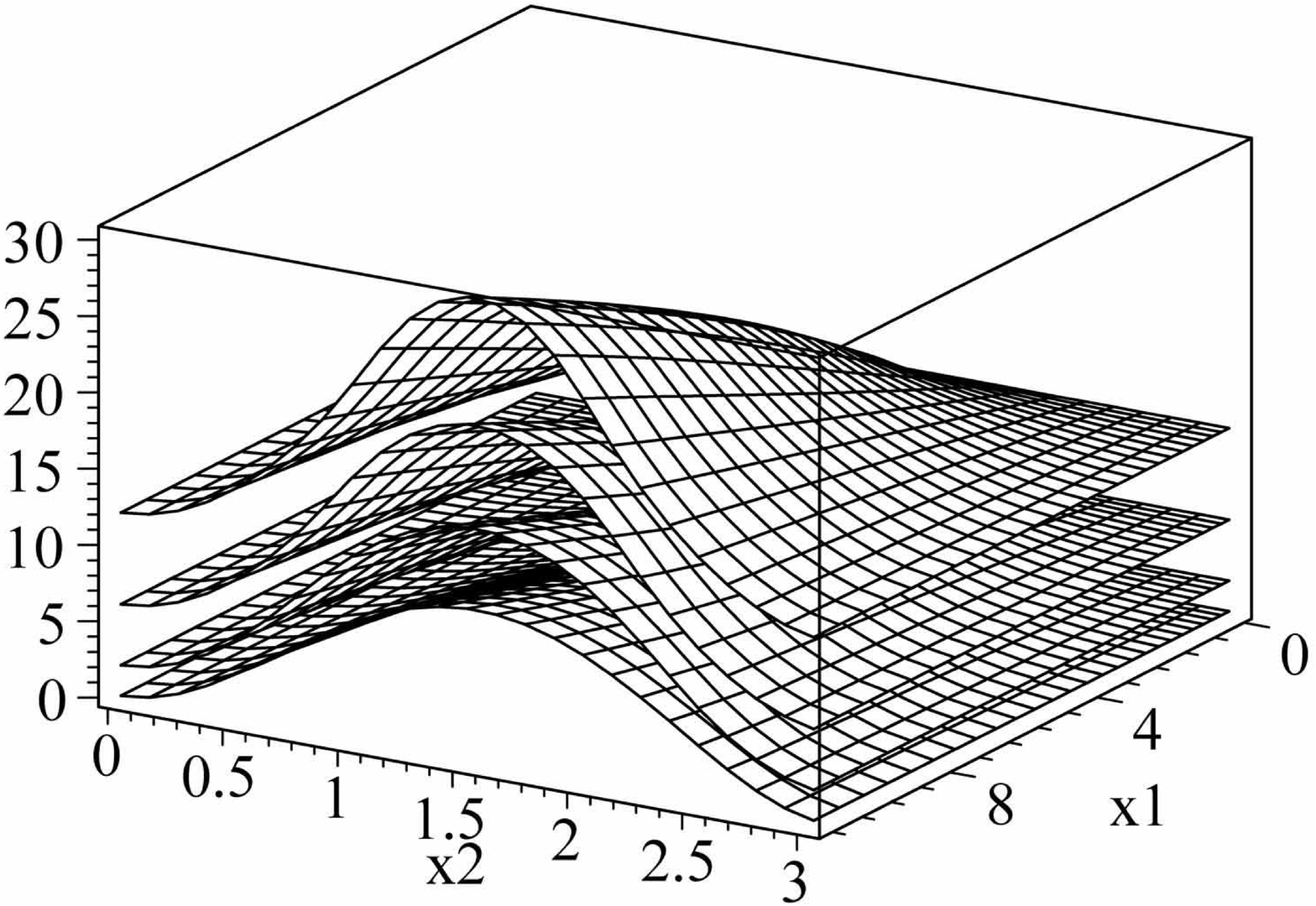}a
\includegraphics[width=0.49\textwidth,height=0.23\textwidth]{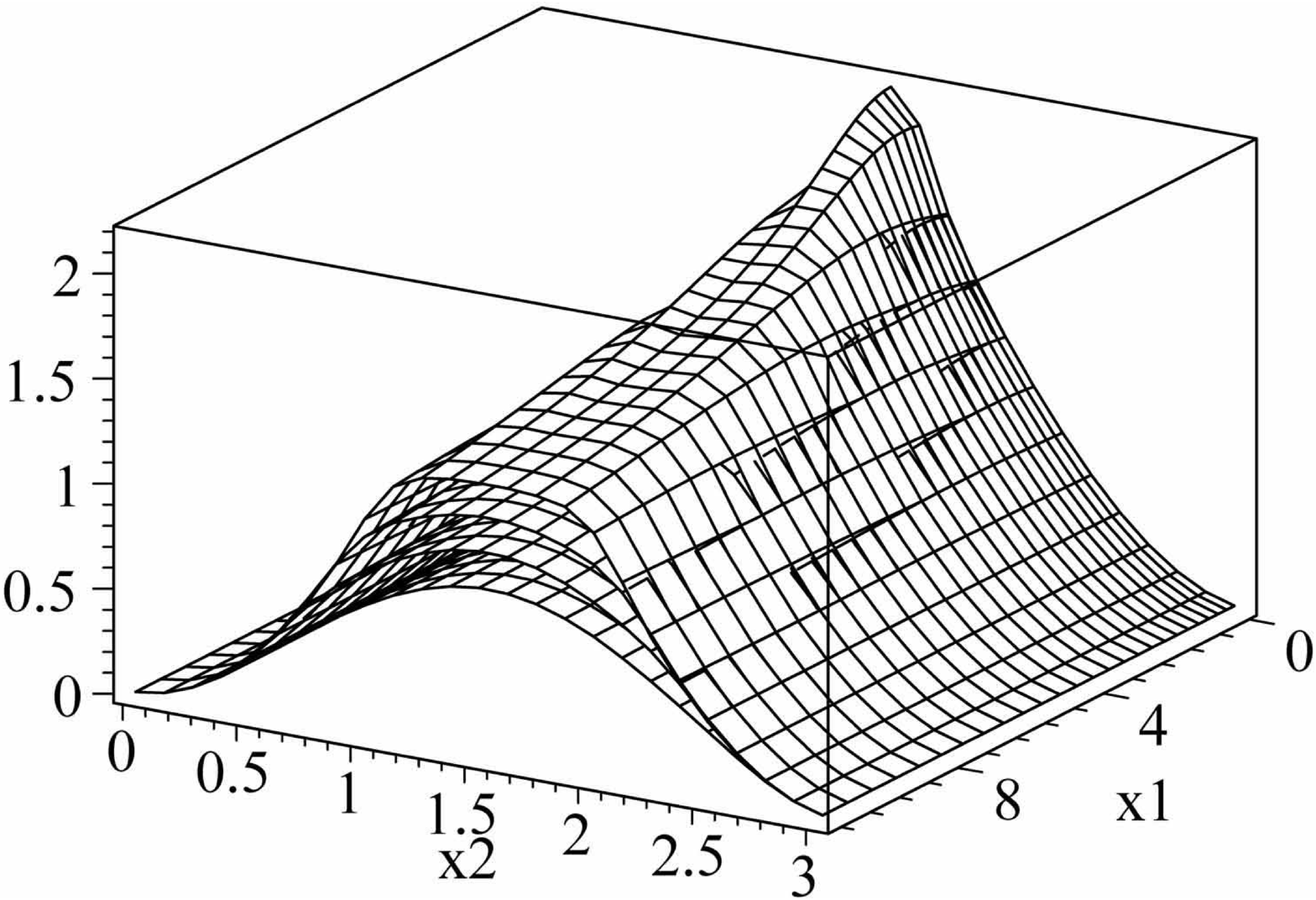}b \\
\includegraphics[width=0.49\textwidth,height=0.23\textwidth]{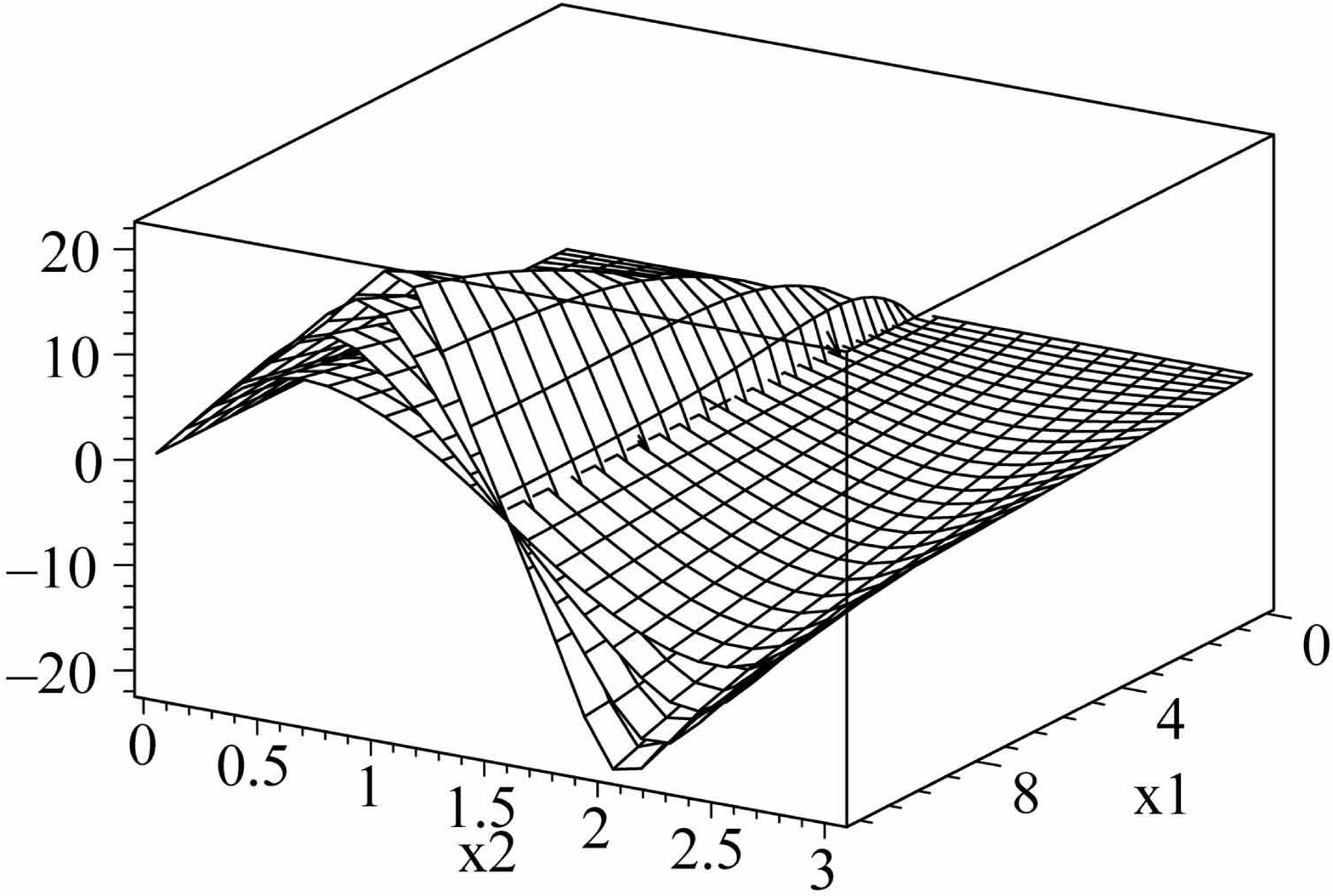}c
\includegraphics[width=0.49\textwidth,height=0.23\textwidth]{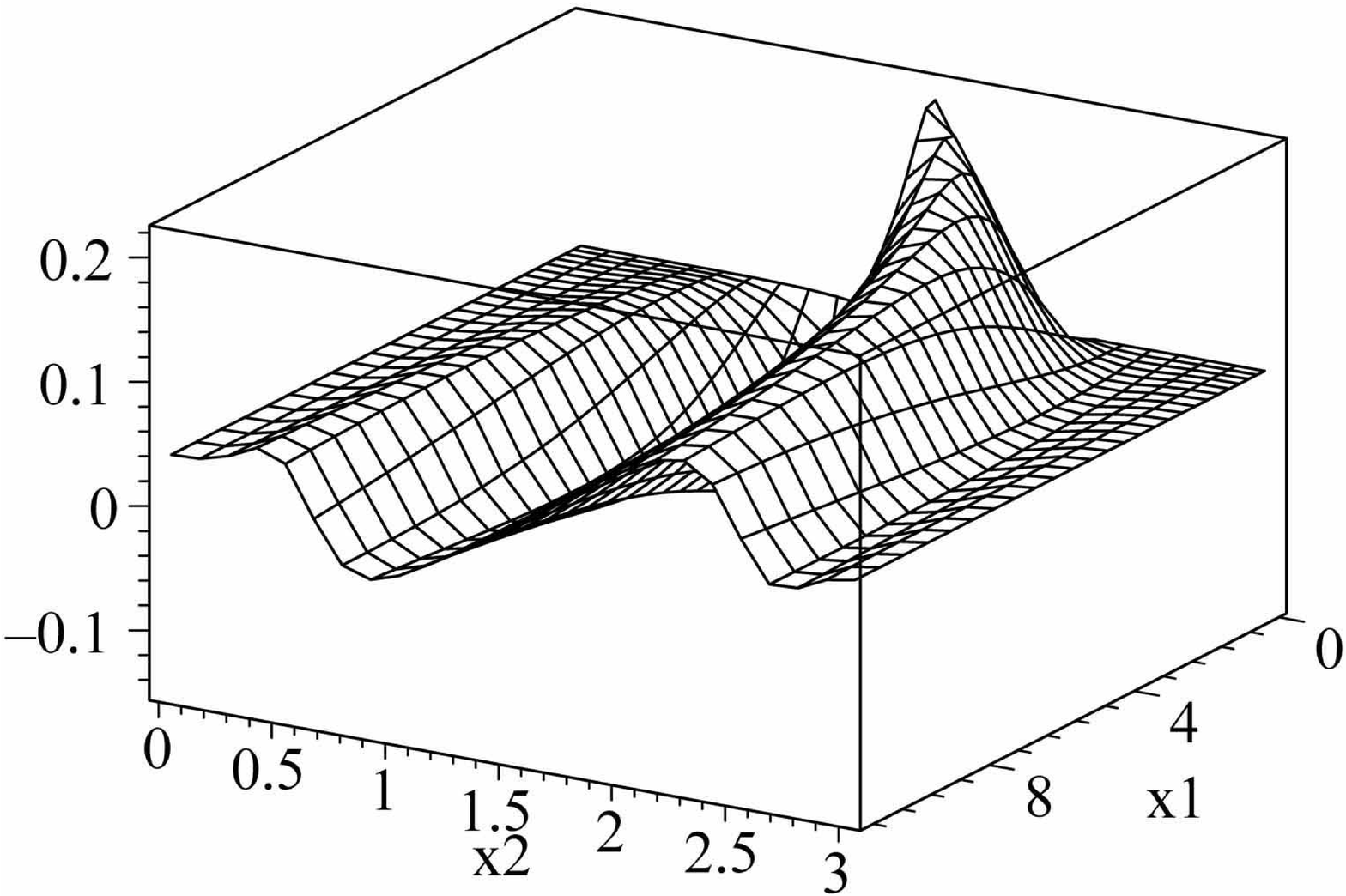}d
\caption{a. The eigenvalues (potentials surfaces)
$E_{i_3}^{(3)}(x_2,x_1)$,  $i_3=1,2,3,4$ of the problem (\ref{mu3}).
b-c.  their parametric derivatives $\partial E_{i_3}^{(3)}(x_2,x_1)/\partial x_1$ and
$\partial E_{i_3}^{(3)}(x_2,x_1)/\partial x_2$, $i_3=1,2,3,4$.
d. their mixed parametric derivative $\partial E_{1}^{(3)}(x_2,x_1)/\partial x_1\partial x_2$}
 \label{e}
\end{figure}

In \textbf{Step 1} the two-parametric problem (\ref{mu3})
with boundary conditions following from (\ref{x5})
is solved for each values of $x_1$ and  $x_2$
with help of the adaptation of ODPEVP program  \cite{ODPEVP},
named here as  ODPEVP 2.0. The eigenvalues (potentials surfaces)
$E_{i_3}^{(3)}(x_2;x_1)\equiv E_{i_3}^{(3)}(x_2,x_1)$ and their
parametric derivatives are presented on Fig. \ref{e}. One can see
from  Fig. \ref{e}, the potential surfaces are symmetric with
respect to axis $x_2=\pi/2$, then the partial derivative
$\partial E_{i_3}^{(3)}(x_2,x_1)/\partial x_2=0$ for $i_3=1,2,...$.

\begin{figure}[t]
\includegraphics[width=0.49\textwidth,height=0.32\textwidth]{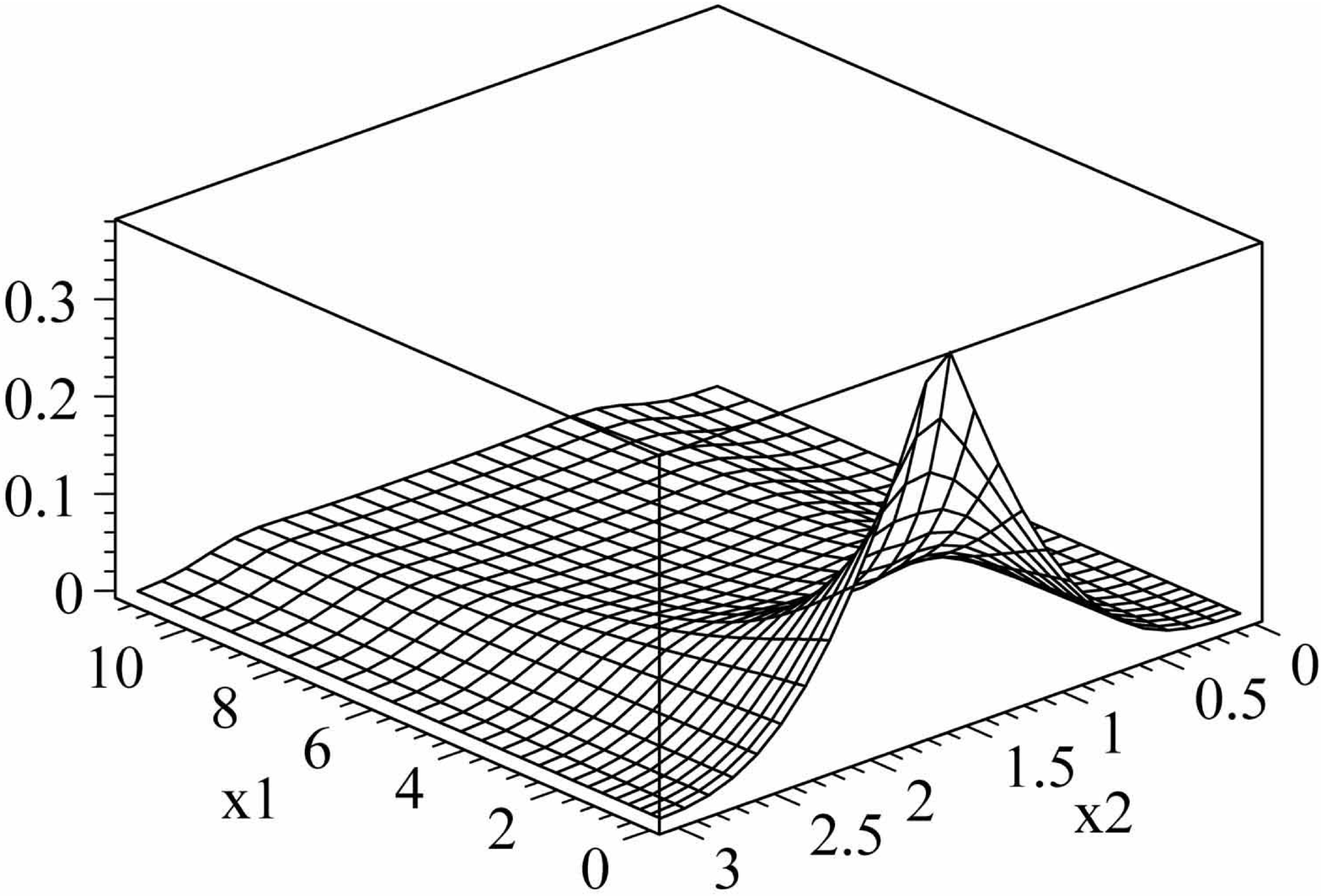}
 \includegraphics[width=0.49\textwidth,height=0.32\textwidth]{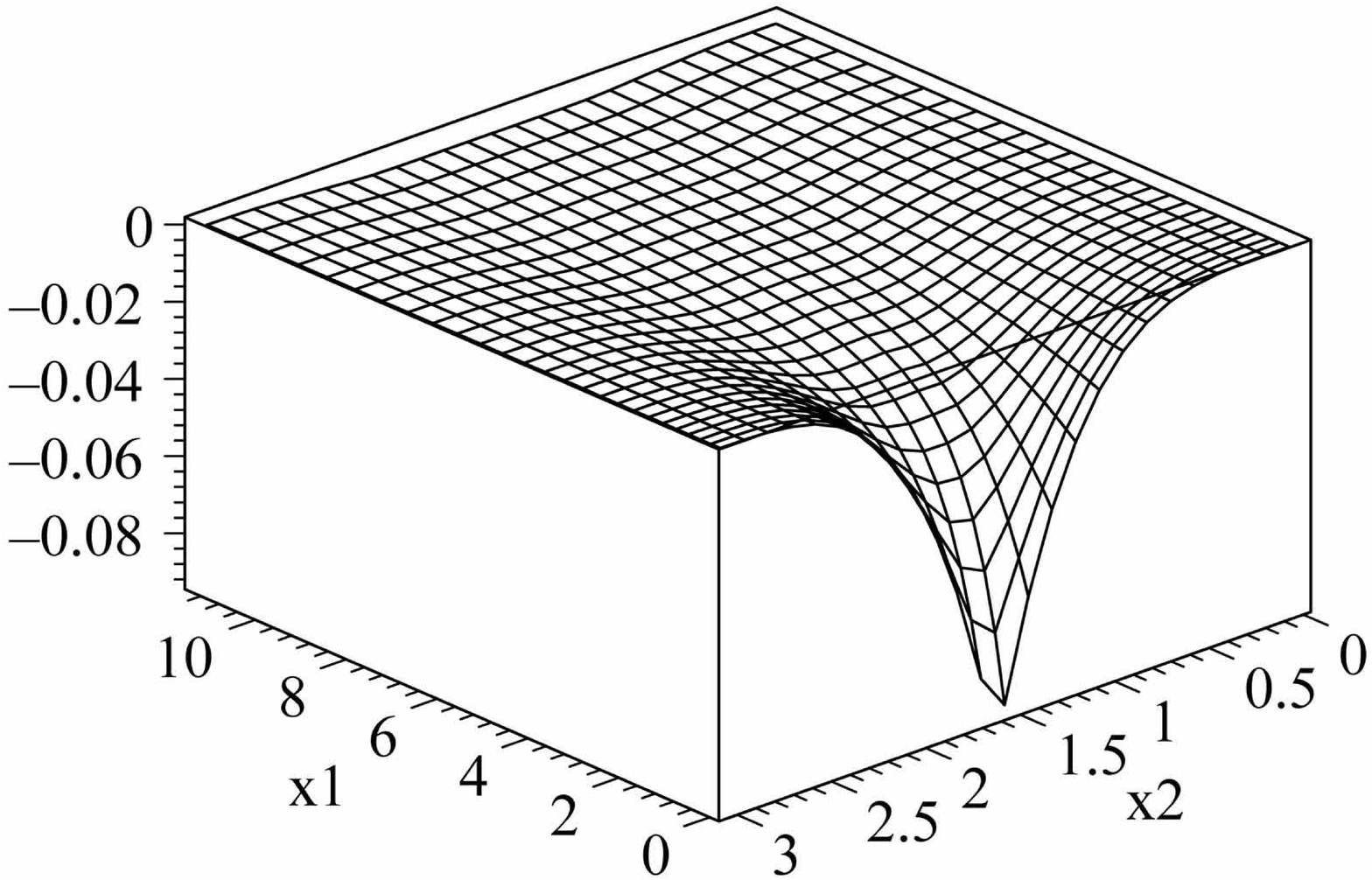}
\caption{Calculated elements $A^{2;00;10}_{12}(x_2;x_1)$ and $A^{2;00;10}_{13}(x_2;x_1)$
of matrix of effective potentials
of Eqs. (\ref{pa5}).}
\label{q}
\end{figure}

\begin{figure}[t]
\includegraphics[width=0.49\textwidth,height=0.32\textwidth]{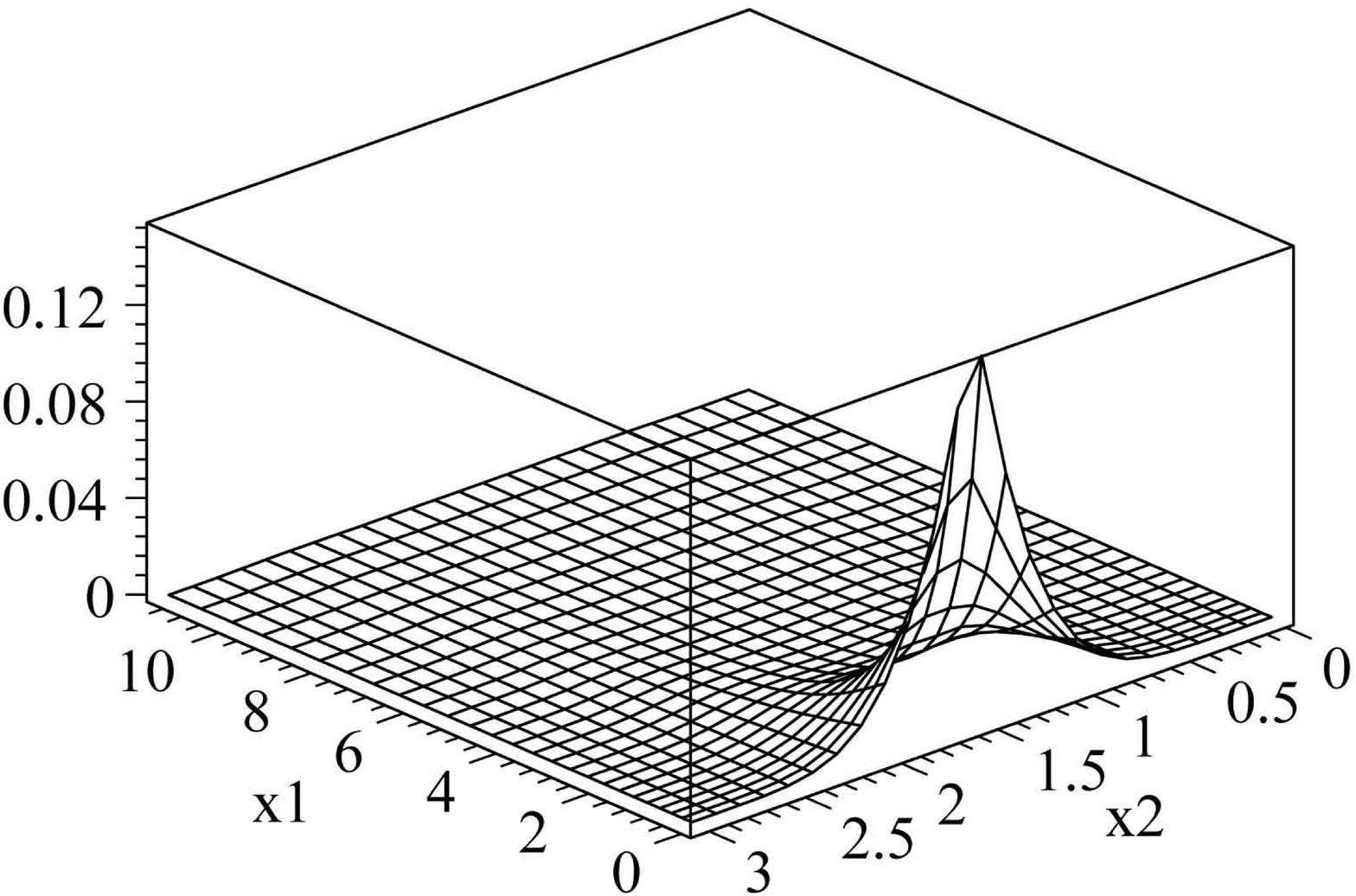}\hfill
\includegraphics[width=0.49\textwidth,height=0.32\textwidth]{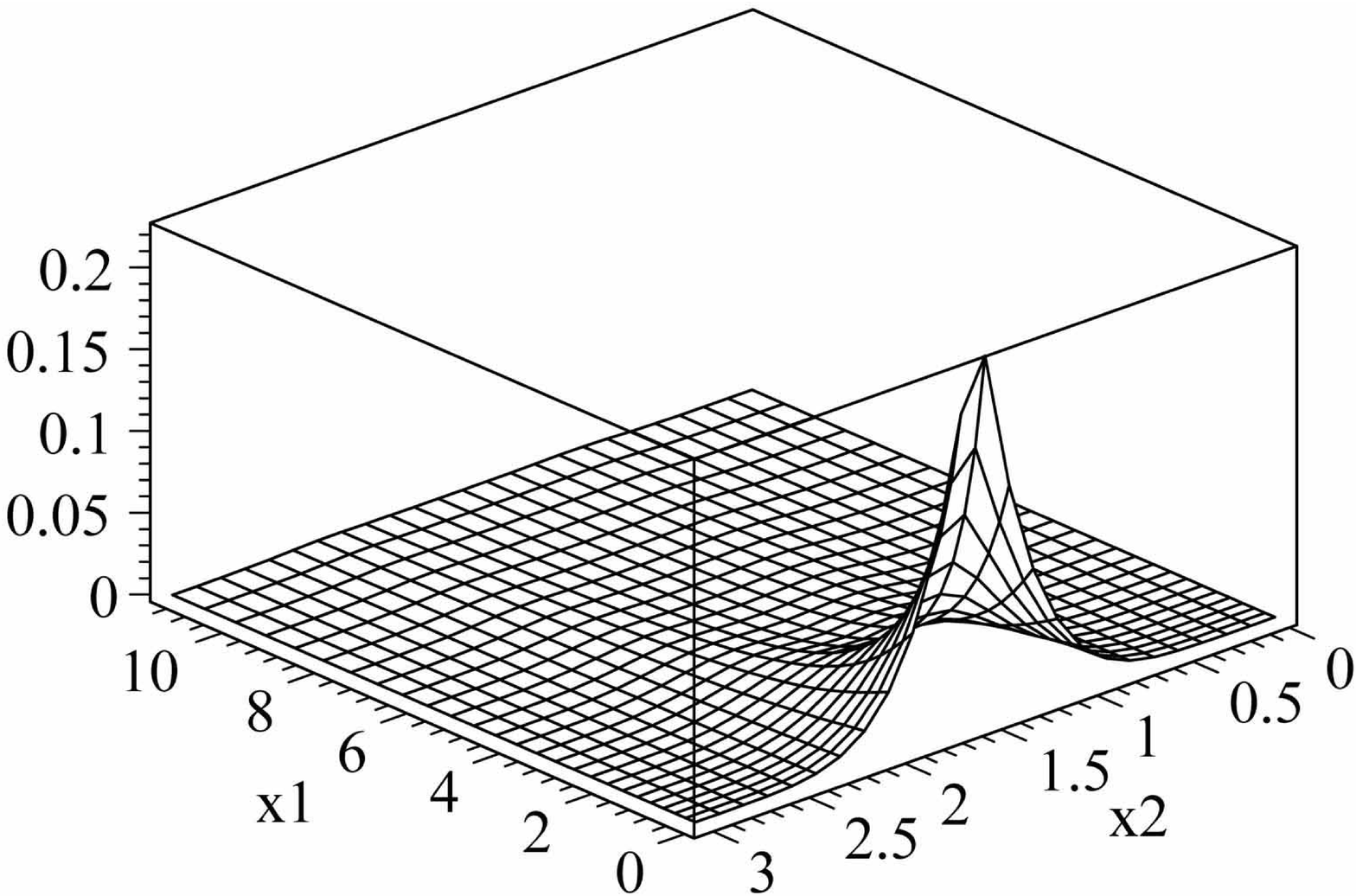}
\caption{Calculated elements $A^{2;10;10}_{11}(x_2;x_1)$
and $A^{2;10;10}_{22}(x_2;x_1)$ of matrix of effective potentials
of Eqs. (\ref{pa5}).}
\label{d}
\end{figure}

\begin{figure}[t]
\includegraphics[width=0.49\textwidth,height=0.32\textwidth]{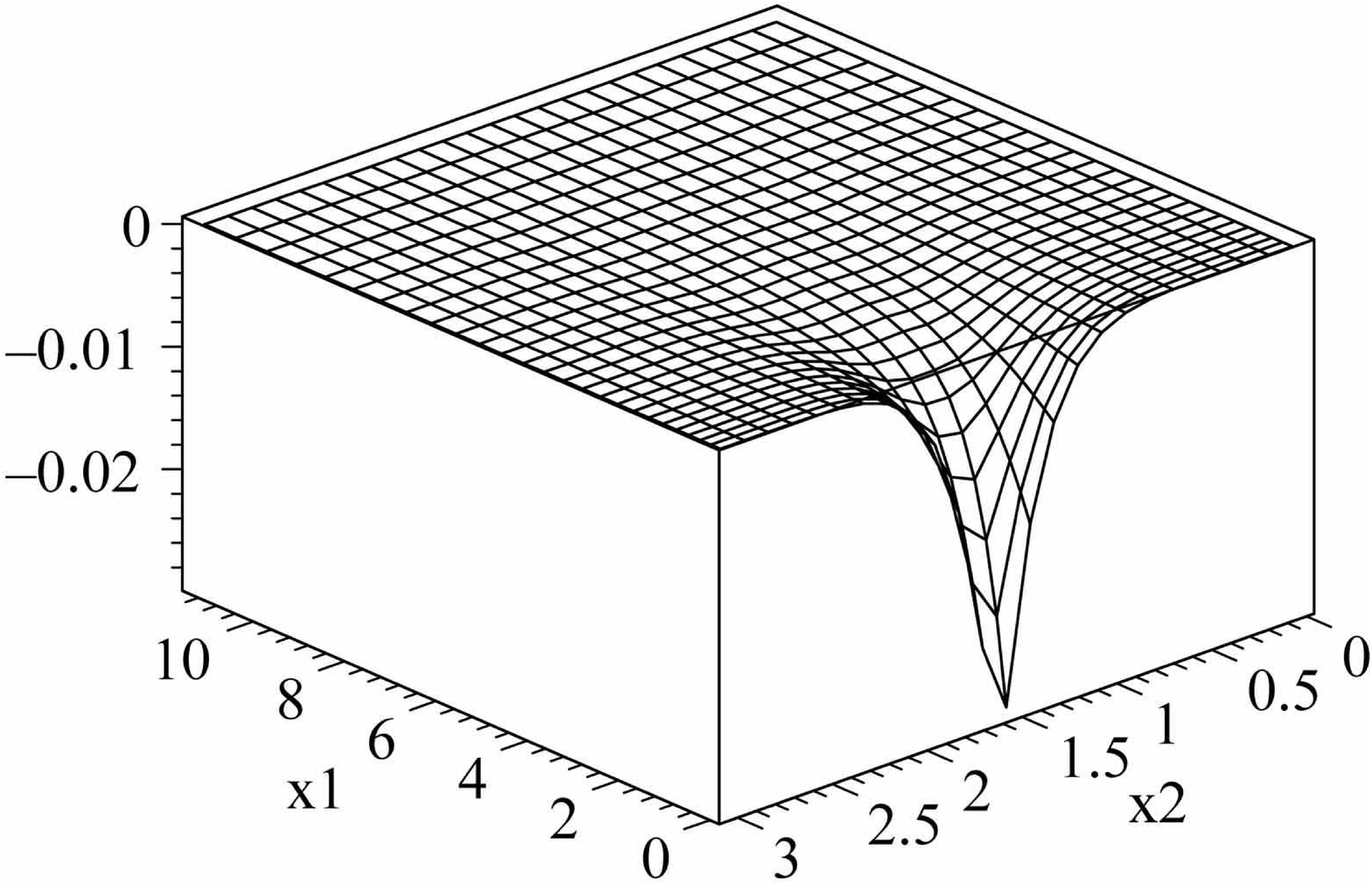}\hfill
\includegraphics[width=0.49\textwidth,height=0.32\textwidth]{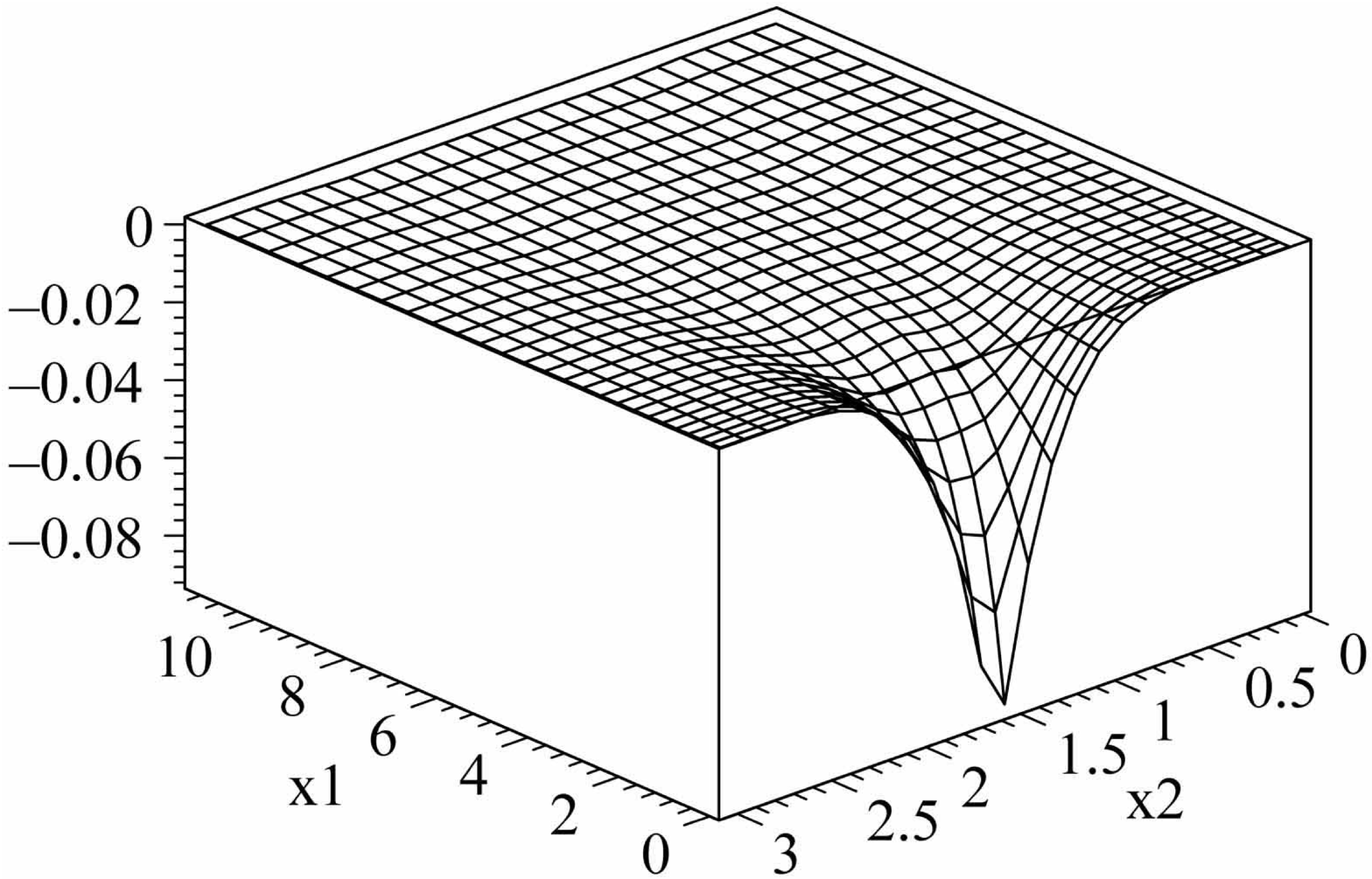}
\caption{Calculated elements $A^{2;10;10}_{12}(x_2;x_1)$
and $A^{2;10;10}_{13}(x_2;x_1)$ of matrix of effective potentials
of Eqs. (\ref{pa5}).}
\label{h}
\end{figure}

In \textbf{Step 2} we find the solution of the   one-parametric problem (\ref{mu2a})
in the series expansion over solutions
of the problem (\ref{mu3}) solved in the \textbf{Step 1}:
\begin{eqnarray}\label{mu23}
\Psi_{i_2}^{(2)}(x_3,x_2;x_1)=\sum_{i_3=1}^{i^{\max}_3}\Psi_{i_3}^{(3)}(x_3;x_2,x_1)
\chi_{i_3i_2}^{(2)}(x_2;x_1).
\end{eqnarray}
Substituting expansion (\ref{mu23}) into equation (\ref{mu2a}) and projecting with
account of orthonormalization conditions (\ref{mu3}) of parametric basis functions
from \textbf{Step~1},
we arrive to the one-parametric problem for unknown vector functions
$\chi_{i_3i_2}^{(2)}(x_2;x_1)$ and eigenvalues (potentials curves) $E_{i_2}^{(2)}(x_1)$:
\begin{eqnarray} \label{pa5}
 \left(
-\frac{1}{\sin^2x_2}\frac{\partial }{\partial x_2}\sin^2x_2\frac{\partial}{\partial x_2}
+\hat V_2(x_2,x_1)
+\frac{E_{i_3}^{(3)}(x_2;x_1)}{2\sin^2x_2}\right.
\\  \nonumber \left.-\frac12E_{i_2}^{(2)}(x_1)\right)\chi_{i_3i_2}^{(2)}(x_2;x_1)+\sum_{j_3=1}^{i^{\max}_3}
 \langle i_3|\Bigl[H_2,j_3\rangle\Bigr] \chi_{j_3i_2}^{(2)}(x_2;x_1)=0,
\\ \nonumber
 \langle i_3|\Bigl[H_2,j_3\rangle\Bigr] =\left(A^{2;10;10}_{i_3j_3}(x_2;x_1)
-\frac{1}{\sin^2x_2}\frac{\partial}{\partial x_2} \sin^2x_2A^{2;00;10}_{i_3j_3}(x_2;x_1)
 \right. \\ \nonumber \left.- A^{2;00;10}_{i_3j_3}(x_2;x_1)
\frac{\partial}{\partial x_2} \right).\end{eqnarray}
with boundary conditions following from (\ref{x5}).
Substituting expansion (\ref{mu23}) into (\ref{mu2ab}), we have orthonormalization conditions
\begin{eqnarray}\label{pa8}
\sum_{j_3=1}^{i^{\max}_3}\int_{{\bf X_2}} \sin^2x_2dx_2\chi_{j_3i_2}^{(2)}(x_2;x_1)
\chi_{j_3j_2}^{(2)}(x_2;x_1)=\delta_{i_2j_2}.
\end{eqnarray}
In (\ref{pa5}) we have definitions of elements of matrix of effective potentials:
{ \begin{eqnarray}
A^{2;l_2l_1;r_2r_1}_{i_3j_3}(x_2;x_1)=
\int_{{\bf X_3}} \sin x_3 dx_3 \frac{\partial^{l_2+l_1}\Psi_{i_3}^{(3)}(x_3;x_2,x_1)}{\partial x_2^{l_2}\partial x_1^{l_1}}
\frac{\partial^{r_2+r_1}\Psi_{j_3}^{(3)}(x_3;x_2,x_1)}{\partial x_2^{r_2}\partial x_1^{r_1}}
,\!\!\!\!\!\!\!\!\!\!\!\!\nonumber\\ \label{a2a}
\frac{\partial^{0}}{\partial x_2^{0}\partial x_1^{0}}\Psi_{i_3}^{(3)}(x_3;x_2,x_1)
\equiv\Psi_{i_3}^{(3)}(x_3;x_2,x_1).\qquad
\end{eqnarray}
}
Note that this problem is similar to problem (\ref{pa1}) from Algorithm 1, but
elements of  matrix of effective potentials are calculated here with eigenfunctions
(\ref{mu3})
and their derivatives by parameters $x_2$, $x_1$ by program ODPEVP 2.0 with accuracy  $O(h^{2p})$,
where $p$ is degree of approximation in a finite element grid  \cite{ODPEVP}.
As an example, some elements of effective potential matrix
are shown in Figs. \ref{q}, \ref{d} and \ref{h}.
The eigenvalues (potential curves) $E_{i_2}^{(2)}(x_1)$  of (\ref{pa5}) calculated here
by program KANTBP 3.0 look as the same as in Figs. \ref{a0} calculated by Algorithm 1.

In \textbf{Step 3} we find the solution of the problem (\ref{mu1a})
in the series expansion over solutions
of problem (\ref{mu2a})--(\ref{mu2ab}) solved in the \textbf{Step 2}:
\begin{eqnarray}\label{mu12}
\Psi_{i_1}^{(1)}(x_3,x_2,x_1)=\sum_{i_2=1}^{i^{\max}_2}
\Psi_{i_2}^{(2)}(x_3,x_2;x_1)\chi_{i_2i_1}^{(1)}(x_1).
\end{eqnarray}
Substituting expansion (\ref{mu12}) into equation (\ref{mu1a}) and projecting with
account of orthonormalization conditions (\ref{mu2ab})
of parametric basis functions from \textbf{Step 2},
we arrive to the problem for unknown vector functions $\chi_{i_2i_1}^{(1)}(x_1)$:
\begin{eqnarray} \label{pa6}
\left(
-\frac{1}{x_1^5}\frac{\partial }{\partial x_1}x_1^5\frac{\partial}{\partial x_1}
+\frac{2E_{i_2}^{(2)}(x_1)-4}{x_1^2}-2E_{i_1}^{(1)}\right)\chi_{i_2i_1}^{(1)}(x_1)
\\ +\sum_{j_2=1}^{i^{\max}_2} \langle i_2|\Bigl[H_1,j_2\rangle\Bigr]
\chi_{j_2i_1}^{(1)}(x_1)=0,\nonumber
\\
 \langle i_2|\Bigl[H_1,j_2\rangle\Bigr] =\left(A^{1;1;1}_{i_2j_2}(x_1)
-\frac{1}{x_1^5}\frac{\partial}{\partial x_1}x_1^5A^{1;0;1}_{i_2j_2}(x_1)
- A^{1;0;1}_{i_2j_2}(x_1)\frac{\partial}{\partial x_1} \right). \nonumber
\end{eqnarray}
with boundary conditions following from (\ref{x5}).
Substituting expansion (\ref{mu12}) into (\ref{mu1ab}), we have required
orthonormation conditions
\begin{eqnarray} \label{pa7}
\sum_{j_2=1}^{i^{\max}_2}\frac{1}{8}\int_{{\bf X_1}} x_1^5dx_1\chi_{j_2i_1}^{(1)}(x_1)
\chi_{j_2j_1}^{(1)}(x_1)=\delta_{i_1j_1}.
\end{eqnarray}
In (\ref{pa6}) we have definitions of elements of matrix of effective potentials:
\begin{eqnarray*}
A^{1;l_1;r_1}_{i_2j_2}(x_1)=
\int_{{\bf X_3}\cup{\bf X_2}}\!\!\! \sin x_3 dx_3\sin^2x_2dx_2
\frac{\partial^{l_1}\Psi_{i_2}^{(2)}(x_3,x_2;x_1)}{\partial x_1^{l_1}}
\frac{\partial^{r_1}\Psi_{j_2}^{(2)}(x_3,x_2;x_1)}{\partial x_1^{r_1}},\\
\frac{\partial^{0}\Psi_{i_2}^{(2)}(x_3,x_2;x_1)}{\partial x_1^{0}}
\equiv\Psi_{i_2}^{(2)}(x_3,x_2;x_1).
\end{eqnarray*}
Substituting expansion (\ref{mu23}), we reduce matrix elements
$A^{1;l_1;r_1}_{i_2j_2}(x_1)$
to integrals by variable $x_2$ only
via matrix elements
$A^{2;l_1l_2;r_1r_2}_{i_3j_3}(x_2;x_1)$
calculated with help of improved parametric basis functions (\ref{mu23})
from \textbf{Step 2}:
\begin{eqnarray}&&\label{pa9}
A^{1;l_1;r_1}_{i_2j_2}(x_1)=
\sum_{i_3,j_3}\sum_{k_l=0}^{l_1}\sum_{k_r=0}^{r_1}\frac{l_1!}{k_l!(l_1-k_l)!}
\frac{r_1!}{k_r!(r_1-k_r)!}
\\&&\times\int_{{\bf X_2}} \sin^2x_2dx_2
A^{2;0k_l;0k_r}_{i_3j_3}(x_2;x_1)
\frac{\partial^{l_1-k_l}\chi_{i_3i_2}^{(2)}(x_2;x_1)}{\partial x_1^{l_1-k_l}}
\frac{\partial^{r_1-k_r}\chi_{j_3j_2}^{(2)}(x_2;x_1)}{\partial x_1^{r_1-k_r}}. \nonumber
\end{eqnarray}
Note that set of Eqs. (\ref{pa6}) is similar to  Eqs. (\ref{pa1}) from Algorithm 1,
and the potential curves $E_{i_2}^{(2)}(x_1)$ and elements
of matrices $A^{1;l_1;r_1}_{i_2j_2}(x_1)$  calculated
by Algorithm 2  look as the same as in Figs. \ref{a0} and \ref{a1} calculated by Algorithm 1.
We can wait that
using expansion (\ref{mu23}) over the two-parametric basis functions (\ref{mu3})
will have a higher rate of convergence and give corresponding elements
of matrices of effective potentials  (\ref{pa9}) with better accuracy in comparison with
 expansion (\ref{mu23a}) over the Legendre polynomials.
However, matrix elements have a more complicate structure and additional
numerical integration by variable $x_3$ performed
in \textbf{Step 2}  is needed in comparison with (\ref{pd}).

\section{Conclusions}
In this paper we presented a symbolic algorithm
for reduction of multistep adiabatic equations,
corresponding to the MultiStep Generalization of Kantorovich Method,
for solving multidimensional boundary-value problems and consider examples
of its application to Helium atom calculation.
Achievement of this approach
consist in facts that, on each step solution subject to boundary
conditions, elements of matrix of effective potentials
calculated with controllable accuracy
have smooth behavior with respect to parameters like that
in spheroidal coordinates  \cite{abramov}.
These facts together with consistency of Kantorovich expansion in a vicinity
triple-collision point $R\to0$
including logarithmic terms corresponding to Fock expansion  \cite{AAPV91,KV87}
provide as shown above a reasonable rate of convergence of these expansions
and upper estimations of energy eigenvalues \cite{M85}
Moreover, asymptotics of these expansions at large value
$R\to\infty$ in vicinities of pair collision points  of limit of separated atom
are compatible with asymptotic states
needed for solving a scattering problem below three body threshold, as shown in papers
\cite{AAPV91,KV87}.

Elaboration final version of Program KANTBP 3.0 for solving the problem with respect to
unknowns (i.e. calculation of improved
parametric basis functions in Algorithm 2) from \textbf{Steps 2--(n-1)}, with matrices of
variable coefficients calculated and presented above is in progress.

Generalization of MultiStep Kantorovich method presented above
reduce to the set of $2N-1$ of multiparametric eigenvalue problem
for set of $\sim 10$ ordinary second-order differential equations
that can solve naturally by each of $N-1$, $N-2$, ..., $1$, $0$
independent parameter using MPI and/or GRID technology that will be
elaborated in our further investigations.

The computational scheme, the SNA, and   the complex of programs allow extension  for the analysis of spectral characteristics of both electron(hole), impurity and excitonic states in  nanoscale quantum-dimensional models like QWs \cite{casc09}, QWrs\cite{JOPA}, QDs \cite{4202} with different geometry of structure and spatial form of confining potential and external fields.

\section*{Acknowledgements}
This work was done within the framework of the Protocol
No.3967-3-6-09/11 of collaboration between JINR and RAU in dynamics
of finite-dimensional models and nanostructures in external fields.
The work was supported by RFBR (grants 10-01-00200, 08-01-00604) and
by the grant No. MK-2344.2010.2 of the President of Russian
Federation.

\bibliographystyle{elsart-num-sort}

\begin{thebibliography}{00}
\bibitem{abramov} D.I. Abramov,
Hyperspherical Coulomb spheroidal
representation in the Coulomb three-body
problem J. Phys. B {41} (2008)  175201.
\bibitem{AAPV91} A.G. Abrashkevich, D.G. Abrashkevich, I.V. Puzynin, S. I. Vinitsky,
Adiabatic hyperspherical representation in barycentric
coordinates for helium-like systems
J. Phys. B  {24} (1991) 1615-1638.
\bibitem{A00} A.G. Abrashkevich, M.S. Kaschiev, S.I. Vinitsky,
A new method for solving an eigenvalue problem for a
system of three Coulomb particles within the
hyperspherical adiabatic representation. J. Comput. Phys., \textbf{163} (2000)  328-348.
\bibitem{Baer06} M. Baer, Beyond Born–Oppenheimer, conical intersections and electronic
nonadiabatic coupling terms. John Wiley \& Sons Inc., Hoboken, 2006.
\bibitem{BH54} M. Born, K. Huang, Dynamical Theory of Crystal Lattices.
Clarendon, Oxford, 1954.
\bibitem{BO27} M. Born,  J.R. Oppenheimer, Zur Quantentheorie der Molekeln.
Annalen der physik 84 (1927) 457.
\bibitem{KANTBP}O. Chuluunbaatar, A.A. Gusev, A.G. Abrashkevich, A. Amaya-Tapia, M.S. Kasc\-hiev, S.Y. Larsen, S.I. Vinitsky,
KANTBP: A program for
computing energy levels, reaction matrix and radial wave
functions in the coupled-channel hyperspherical adia-batic
approach. Comput. Phys. Commun. {  177} (2007) 649--675.
\bibitem{ODPEVP}  O. Chuluunbaatar, A.A. Gusev, S.I. Vinitsky, A.G. Abrashkevich,  ODPEVP: A program for computing
eigenvalues and eigenfunctions and their first derivatives with
respect to the parameter of the parametric self-adjoined
Sturm-Liouville problem. Comput. Phys. Commun.  {180} (2009) 1358--1375.
\bibitem{D94} G.W.F. Drake, Zong-Chao Van, Variational eigenvalues for the S states of helium
Chem. Phys. Lett. {\bf 229} (1994) 486--490.
\bibitem{DMV87} V.M. Dubovik, B.L. Markovski, S.I. Vinitsky,
Multistep adiabatic approximation, preprint JINR E4-87-743, Dubna, 1987;
%retrieved  from
http://www-lib.kek.jp/cgi-bin/img\_index?8801189.
\bibitem{G98} J.J. De Groote, M. Masili, J.E. Hornos,
Highly excited states for the helium atom in the
hyperspherical adiabatic approach.
J. Phys. B  {\bf 31} (1998) 4755--4764.
\bibitem{4202} A.A. Gusev, O. Chuluunbaatar, V.P. Gerdt, V.A. Rostovtsev, S.I. Vinitsky, V.L. Derbov, V.V. Serov, Symbolic-Numeric Algorithms for Computer Analysis of   Spheroidal Quantum Dot Models. in Proc. of The 12th International Workshop on Computer Algebra in Scientific Computing (CASC 2010) Tsakhkadzor, Armenia September 5 - 12, 2010 (to appear); http://arxiv.org/abs/1004.4202.
    \bibitem{KV87} M.B. Kadomtsev, S.I. Vinitsky,
Adiabatic representation for the three-body problem in
hyperspherical coordinates: I. Statement of the problem
J. Phys. B   {20} (1987) 5723-5736.
\bibitem{KK64} L.V. Kantorovich, V.I. Krylov,
Approximate Methods of Higher Analysis. Wiley, New York, 1964.
\bibitem{MV89} Topological phases in quantum theory.
Eds. B. Markovski, S.I. Vinitsky World Sci., Singapore,  1989.
\bibitem{M85} J. Makarewicz, Adiabatic multi-step separation method and its application
to coupled oscillators. Theor. Chim. Acta {68} (1985) 321--334.
\bibitem{casc09} S.I. Vinitsky, O. Chuluunbaatar, V.P. Gerdt, A.A. Gusev and V.A. Rostovtsev Symbolic-Numerical Algorithms for Solving Parabolic Quantum Well Problem with Hydrogen-Like Impurity   	Lect. Notes in Computer Science, 5743, 334-349 (2009).
\bibitem{JOPA} O. Chuluunbaatar, A.A. Gusev, V.L. Derbov, M.S. Kaschiev, L.A. Melnikov, V.V. Serov and S.I. Vinitsky, Calculation of a hydrogen atom photoionization in a strong magnetic field by using the angular oblate spheroidal functions, J. Phys. A: Mathematical and Theoretical 40,	11485-11524 (2007).
\end{thebibliography}

 \end{document}